\documentclass[a4paper,clock]{article}
\usepackage[dvips]{epsfig}
\usepackage{amssymb}
\usepackage{bm}
\usepackage{dcolumn}
\catcode`\@=11
\def\marginnote#1{}
\newcount\hour
\newcount\minute
\newtoks\amorpm
\hour=\time\divide\hour by60
\minute=\time{\multiply\hour by60 \global\advance\minute
by-\hour}\edef\standardtime{{\ifnum\hour<12
\global\amorpm={am}%
        \else\global\amorpm={pm}\advance\hour by-12 \fi
        \ifnum\hour=0 \hour=12 \fi
        \number\hour:\ifnum\minute<10
0\fi\number\minute\the\amorpm}}
\edef\militarytime{\number\hour:\ifnum\minute<10
0\fi\number\minute}

\def\draftlabel#1{{\@bsphack\if@filesw {\let\thepage\relax
   \xdef\@gtempa{\write\@auxout{\string
      \newlabel{#1}{{\@currentlabel}{\thepage}}}}}\@gtempa
   \if@nobreak \ifvmode\nobreak\fi\fi\fi\@esphack}
        \gdef\@eqnlabel{#1}}
\def\@eqnlabel{}
\def\@vacuum{}
\def\draftmarginnote#1{\marginpar{\raggedright\scriptsize\tt#1}}
\def\draft{\oddsidemargin -.5truein
        \def\@oddfoot{\sl preliminary draft \hfil
        \rm\thepage\hfil\sl\today\quad\militarytime}
        \let\@evenfoot\@oddfoot \overfullrule 3pt
        \let\label=\draftlabel
        \let\marginnote=\draftmarginnote

\def\@eqnnum{(\theequation)\rlap{\kern\marginparsep\tt\@eqnlabel}%
\global\let\@eqnlabel\@vacuum}  }

\def\numberbysection{\@addtoreset{equation}{section}
        \def\theequation{\thesection.\arabic{equation}}}

\def\underline#1{\relax\ifmmode\@@underline#1\else
 $\@@underline{\hbox{#1}}$\relax\fi}

\catcode`@=12
\relax

\numberbysection
\advance \topmargin by -\headheight
\advance \topmargin by -\headsep

\evensidemargin \oddsidemargin
\def\br{\begin{eqnarray}}
\def\er{\end{eqnarray}}
\def\be{\begin{equation}}
\def\ee{\end{equation}}

\def\({\left(}
\def\){\right)}

\relax

\def\a{\alpha}

\def\l{\lambda}

\def\pa{\partial}

\def\tp0{\Theta_{+}^{(0)}}
\def\tm0{\Theta_{-}^{(0)}}

\def\f#1#2#3 {f^{#1#2}_{#3}}

\def\win1{{\sf w_{1+\infty}}}

\def\Win1{{\sf W_{1+\infty}}}

\def\rlx{\relax\leavevmode}
\def\inbar{\vrule height1.5ex width.4pt depth0pt}
\def\IZ{\rlx\hbox{\sf Z\kern-.4em Z}}
\def\IR{\rlx\hbox{\rm I\kern-.18em R}}
\def\IC{\rlx\hbox{\,$\inbar\kern-.3em{\rm C}$}}
\def\IN{\rlx\hbox{\rm I\kern-.18em N}}
\def\IO{\rlx\hbox{\,$\inbar\kern-.3em{\rm O}$}}
\def\IP{\rlx\hbox{\rm I\kern-.18em P}}
\def\IQ{\rlx\hbox{\,$\inbar\kern-.3em{\rm Q}$}}
\def\IF{\rlx\hbox{\rm I\kern-.18em F}}
\def\IG{\rlx\hbox{\,$\inbar\kern-.3em{\rm G}$}}
\def\IH{\rlx\hbox{\rm I\kern-.18em H}}
\def\II{\rlx\hbox{\rm I\kern-.18em I}}
\def\IK{\rlx\hbox{\rm I\kern-.18em K}}
\def\IL{\rlx\hbox{\rm I\kern-.18em L}}
\def\one{\hbox{{1}\kern-.25em\hbox{l}}}
\def\0#1{\relax\ifmmode\mathaccent"7017{#1}%
B        \else\accent23#1\relax\fi}

\def\RQE#1#2#3{{\sl Radiophysics and Quantum Electronics} {\bf #1}, #2, #3}
\def\PRL#1#2#3{{\sl Phys. Rev. Lett.} {\bf#1}, #2, #3}
\def\NPB#1#2#3{{\sl Nucl. Phys.} {\bf #1}, #2B, #3}

\def\CMP#1#2#3{{\sl Commun. Math. Phys.} {\bf #1}, #2, #3}

\def\PRE#1#2#3{{\sl Phys. Rev.} {\bf #1}, #2E, #3}

\def\PLA#1#2#3{{\sl Phys. Lett.} {\bf #1}, #2A, #3}

\def\PLB#1#2#3{{\sl Phys. Lett.} {\bf #1}, #2B, #3}

\def\PTP#1#2#3{{\sl Prog. Theor. Phys.} {\bf #1}, #2, #3}

\def\IJMPB#1#2#3{{\sl Int. J. Mod. Phys.} {\bf #1}, #2B, #3}

\def\JPA#1#2#3{{\sl J. Physics} {\bf A#1}, #2, #3}

\def\JHEP#1#2#3{{\sl JHEP} {\bf #1}, #2, #3}

\def\SAM#1#2#3{{\sl Stud. Appl. Math.} {\bf #1},#2, #3}

\def\Nonl#1#2#3{{\sl Nonlinearity} {\bf #1}, #2, #3}

\def\JPAMG#1#2#3{{\sl J. Physics A: Math. Gen.} {\bf #1}, #2A, #3}
\def\JGP#1#2#3{{\sl Journal of Geometry and Physics} {\bf #1}, #2, #3}
\def\EPL#1#2#3{{\sl Europhysics Letters} {\bf #1}, #2, #3}
\def\ScR#1#2#3{{\sl Sci. Rep.} {\bf #1},#2, #3}

\def\JPCS#1#2#3{{\sl J. Phys.: Conf. Ser} {\bf #1},#2, #3}

\def\BJP#1#2#3{{\sl Brazilian Journal of Physics} {\bf #1}, #2, #3}
\def\NJP#1#2#3{{\sl New J. Phys.} {\bf #1}, #2, #3}
\def\DCDS#1#2#3{{\sl Discrete and Continuous Dynamical Systems - S} {\bf #1} (#2) #3} 
\def\CNSNS#1#2#3{{\sl Commun Nonlinear Sci Num Simulat} {\bf #1}, #2, #3} 
\def\JMCh#1#2#3{{\sl J Math Chem} {\bf #1}, #2, #3}

\begin{document}

\begin{titlepage}

\vspace{.2in}
\begin{center}
{\large\bf Riccati-type pseudo-potential approach to quasi-integrability of deformed soliton theories}
\end{center}

\vspace{.2in}

\begin{center}

Harold Blas 
\par \vskip .2in \noindent

$^{(a)}$Instituto de F\'{\i}sica\\
Universidade Federal de Mato Grosso\\
Av. Fernando Correa, $N^{0}$ \, 2367\\
Bairro Boa Esperan\c ca, Cep 78060-900, Cuiab\'a - MT - Brazil. il

\normalsize
\end{center}

\vspace{.3in}

\begin{abstract}
\vspace{.3in}

This review paper explores the Riccati-type pseudo-potential formulation applied to the quasi-integrable sine-Gordon, KdV, and NLS models. The proposed framework provides a unified methodology for analyzing quasi-integrability properties across various integrable systems, including deformations of the sine-Gordon, Bullough–Dodd, Toda, KdV, pKdV, NLS and SUSY sine-Gordon models. Key findings include the emergence of infinite towers of anomalous conservation laws within the Riccati-type approach and the identification of exact non-local conservation laws in the linear formulations of deformed models. As modified integrable models play a crucial role in diverse fields of nonlinear physics—such as Bose–Einstein condensation, superconductivity, gravity models, optics and soliton turbulence—these results may have far-reaching applications.

\end{abstract}

\end{titlepage}

\section{Introduction}

Certain nonlinear field theory models that have significant physical applications and describe solitary waves are not inherently integrable. Recently, the concept of quasi-integrability has been introduced for specific deformations of integrable models. Within this framework, various properties of deformed soliton models—such as the sine-Gordon (SG), nonlinear Schr\"{o}dinger (NLS), Toda, Korteweg-de Vries (KdV), Bullough-Dodd, and SUSY-SG—have been explored using the anomalous zero-curvature formulation \cite{jhep1}-\cite{epl}, as well as the deformed Riccati-type pseudo-potential approach \cite{npb1, jhep33, proc2, paper1, paper2}. For previous findings on certain nonlinear field theories involving solitary waves and their collision dynamics, see, for example, \cite{hietarinta} and the references therein.

Since the early development of integrable models, the existence of an infinite number of conservation laws has been one of the defining characteristics of integrability \cite{das, faddeev1, abdalla, babelon11, babelon21}. The discovery of anomalous conservation laws even in the $N$-soliton sector of integrable models represents a novel aspect of these systems \cite{npb1, intech}. The anomalous charges in deformations of the SG, NLS and KdV models have also been studied using the Riccati-type pseudo-potential approach \cite{npb1, jhep33, proc2, paper2}, reproducing the results obtained via the zero-curvature method \cite{jhep1, npb}. Notably, both the anomalous zero-curvature formulation and the Riccati-type pseudo-potential approach produce anomalous charges that have the same form as those in standard integrable models. Furthermore, the latter approach also generates additional types of anomalous charges.

Standard integrable models such as the SG, KdV, and NLS equations can be derived as special cases of the AKNS system. Consequently, an appropriate deformation of the conventional pseudo-potential approach within AKNS integrable field theory enables the definition of related quasi-integrable models. In \cite{nucci1,nucci2}, both Lax equations and B\"acklund transformations for well-known nonlinear evolution equations were generated using the concept of pseudo-potentials and the properties of the Riccati equation. These methods have been applied across various integrable systems, facilitating the Lax pair formulation, the construction of conservation laws, and the derivation of B\"acklund transformations \cite{nucci1,nucci2, prl1,wadati1}.

In this review, we present a unified framework for analyzing quasi-integrable sine-Gordon (SG), KdV, and NLS models by extending the Riccati pseudo-potential formalism of integrable systems to their quasi-integrable deformations. The Riccati-type pseudo-potential method has been applied to these quasi-integrable models within the deformed AKNS framework \cite{npb1, jhep33, proc2, paper2}. The newly identified properties have been explored in deformations of both the relativistic SG model with topological solitons and the non-relativistic KdV model with non-topological, unidirectional solitons, both conventionally defined for real scalar fields. Additionally, these properties have been examined in NLS-type models, which involve a complex field with envelope solitons. 

We present a critical review of the Riccati-type pseudo-potential framework as developed in \cite{npb1, jhep33, paper2, bjp1}, refining several of the original arguments and providing more precise formulations of key results. We show that the formalism can be systematically understood as specific deformations of the AKNS hierarchy, achieved through an appropriate choice of the coefficients in the Riccati-type system and the introduction of a suitable set of auxiliary fields, as well as a quantity which encodes the deformation of the interaction potential of the model. Furthermore, by revisiting the problem of deformed KdV, we derive a more streamlined and conceptually transparent proof of a general deformation of the non-linear and dispersive terms in the Riccati-type approach. This approach encompasses and extends previous constructions formulated within the anomalous zero-curvature and Hamiltonian frameworks \cite{npb, abhinav2}. 
 
Some of these quasi-integrability properties were also explored in deformed SG, NLS, KdV and potential KdV (pKdV) models in \cite{npb, paper1, proc3, jhep33,proc2, bjp1} by directly constructing novel quasi-conservation laws from the corresponding equations of motion. These studies demonstrated that, despite the loss of full integrability due to deformation, the models retain an infinite hierarchy of anomalous conservation laws. Furthermore, the findings highlighted the role of non-local conserved charges and the persistence of soliton-like structures, reinforcing the broader applicability of quasi-integrability in deformed integrable models.

Within the pseudo-potential framework, a linear system is proposed whose compatibility condition yields the modified AKNS (MAKNS) equations of motion. Quasi-integrable models explored in the literature \cite{jhep1, jhep2, jhep6, jhep3, npb, jhep4, proc1, jhep5, npb1, jhep33, proc2, bjp1} exhibit key structures, including infinite sets of non-local conserved charges and certain linear formulations. Furthermore, in the Riccati-type pseudo-potential approach, deformed SG, NLS, and KdV models have been shown to arise as compatibility conditions of specific linear systems and to possess infinite towers of exact non-local conservation laws.
    
This paper is structured as follows. The next section introduces a deformation of the Riccati system within the AKNS formalism. Section \ref{sec:dsg} explores the deformation of the SG model, its dual formulation, and the linear system approach leading to non-local charges. Section \ref{sec:dnls} discusses the $sl(2)$ deformed AKNS system and its reduction to the modified NLS model. Section \ref{sec:dkdv} examines the modified KdV model and reviews two types of deformations studied in the literature. Finally, section \ref{sec:discuss} presents conclusions and discussions.

\section{Riccati-type pseudo-potentials and modified integrable models}
\label{sec:riccati}

A general scheme to define a quasi-integrable nonlinear partial 
differential equation in the pseudo-potential approach follows by deforming the system of Riccati equations associated to the relevant integrable system. In fact, the AKNS approach to define an integrable system considers the integrability condition for the system of Riccati equations
\br
\label{riccati11}
\pa_{\xi} u &=& A_0 + A_1 u + A_2 u^2,\\
\pa_{\eta} u &=& B_0 + B_1 u + B_2 u^2,\label{riccati22}
\er
where the complex-valued function $u(\xi,\eta, \l)$ is the so-called Riccati pseudo-potential and depends on a spectral parameter $\l$ which is assumed to be independent of the $\xi$ and $\eta$ variables, $\(\xi, \eta\) \in \IR^2$. The partial derivatives with respect to $\xi$ and $\eta$ are denoted by $\pa_{\xi} u$ and $\pa_{\eta} u$, respectively. The coefficients $A_j(\xi, \eta, \l)$ and $B_j(\xi, \eta, \l)$, on the other hand, do depend on $\l$ and are functionals of the field of the specific integrable model. The parameter $\l$ is assumed to be an arbitrary complex constant, i.e. $\l \in \IC$. So, we will regard $u$ as a holomorphic function of $\l$ in such a way that locally one can Taylor expand it in powers of $\l$. The Riccati equation exhibits some interesting mathematical properties \cite {hille} and it has an unique character in the theory of integrable systems (see e.g. \cite{nucci1,nucci2, nucci3}. Moreover, it has important applications in quantum mechanics (see. e.g. \cite{sever, ma, dong}). 

Next, the compatibility condition $(\pa_{\eta} \pa_{\xi} u - \pa_{\xi} \pa_{\eta} u)=0$ provides the equation of motion of the integrable system \cite{prl1, wadati1}. In fact, specific finite polynomial expansions of $A_j$ and $B_j$ in powers of $\l$ provide the equations of interest, such as the Korteweg-de Vries (KdV), sine-Gordon(SG), and nonlinear Schr\"{o}dinger equations (NLS).  

In order to achieve a deformed model one proceeds by incorporating the relevant deformations in Equation (\ref{riccati22}). Specifically, the quantities $B_{0}(\xi, \eta, \lambda)$ and $B_{1}(\xi, \eta, \lambda)$ encode the deformation fields, along with certain auxiliary fields $r_j(\xi, \eta, \lambda)$. The auxiliary fields satisfy specific first-order linear partial differential equations in the $\xi$ variable
\br
\label{aux11}
F_{k}(r_j,u,\pa_{\xi} r_j, X) = 0,\,\,\, k=1,2,..n_1; \,\,\,j=1,2,..n_2,
\er
where the $F_k$'s for $k=1,2,...n_1$, are some functions of $r_j(\xi, \eta, \lambda) , u(\xi, \eta, \lambda) ,\pa_{\xi} r_j(\xi, \eta, \lambda),$ and $X(\xi, \eta)$. The number of equations ($n_1$) and auxiliary fields ($n_2$) in (\ref{aux11}) will be explicitly formulated for each deformed model. The $\lambda$ independent quantity $X$ encodes the deformation away from the relevant integrable model. This function includes both the interaction potential of the deformed model and the associated deformation parameters. These parameters are not required to be small, as the framework does not rely on a small perturbation theory. Instead, the development allows for deformation parameters of arbitrary magnitude.     

In the undeformed integrable case, setting $X=0$ leads to $r_j=0$ and the trivial equations $F_k =0$. As a result, System (\ref{riccati11})-(\ref{riccati22}) simplifies to the standard Riccati equations associated with the corresponding integrable system. 

The compatibility  
condition of the Riccati-type System (\ref{riccati11})-(\ref{riccati22}) can be written as 
\br
\label{compati1}
(\pa_{\eta} \pa_{\xi} u - \pa_{\xi} \pa_{\eta} u)=0.
\er 
Using successively System (\ref{riccati11})-(\ref{riccati22}) one can construct the next two terms entering (\ref{compati1})
\br
\pa_{\eta} \pa_{\xi} u &=& \pa_{\eta}A_0 + A_1 B_0 + (2 A_2 B_0 + A_1 B_1 + \pa_{\eta}A_1) u + (2 A_2 B_1 + A_1 B_2 + \pa_{\eta}A_2 ) u^2 + 2 A_2 B_2  u^3,
\er  
and 
\br
\pa_{\xi} \pa_{\eta} u &=& \pa_{\xi}B_0 + A_0 B_1 + (2 A_0 B_2 + A_1 B_1 + \pa_{\xi}B_1) u + (2 A_1 B_2 + A_2 B_1) u^2+ \pa_{\xi}B_2 + 2 A_2 B_2  u^3.
\er
So, subtructing  the above expressions and substituting into (\ref{compati1}) one gets the equation 
\br
\nonumber
A_1 B_0 -A_0 B_1 + \pa_{\eta} A_0 - \pa_{\xi} B_0 + u [2A_2 B_0 -2 A_0 B_2 + \pa_{\eta} A_1 - \pa_{\xi} B_1] &+& \\
u^2 [A_2 B_1 -A_1 B_2 + \pa_{\eta} A_2 - \pa_{\xi} B_2] &=& 0. \label{AB11}
\er
Furthermore, from (\ref{AB11}) the relevant modified equation of motion is obtained, provided that the equations for the auxiliary fields (\ref{aux11}) are considered.

The introduction of the deformation fields  $r_j$ and $X$ constitutes a foundational aspect of our framework for modifying integrable systems. The auxiliary relations given in Eq. (\ref{aux11}), satisfied by the fields $r_j$, guarantee the consistency of  the deformed equations of motion via the compatibility condition expressed in Eq. (\ref{compati1}). The field $X$ incorporates explicit deformation parameters and encapsulates the modified potential structure of the system, as demonstrated below for the sine-Gordon (SG) and nonlinear Schr\"{o}dinger (NLS) models \cite{npb1, paper2}. In the context of the deformed Korteweg–de Vries (KdV) model, the field $X$ additionally encodes deformations in both the nonlinear and dispersive sectors of the equation \cite{jhep33}. For undeformed, integrable limits of the theory, the deformation field $X$ identically vanishes, which in turn implies that Eq. (\ref{aux11}) admits only trivial solutions $r_j =0$.

An alternative deformation scheme is based on the anomalous zero-curvature framework, wherein the conventional Lax pair $\{L_{1}, L_{2}\}$ is modified through deformations of the terms containing the nonlinear and dispersive terms  \cite{jhep1, jhep3, npb}. This deformation gives rise to an anomalous zero-curvature condition characterized by the emergence of an extra term that breaks the standard flatness condition. Through an Abelianization procedure, this anomalous structure yields an infinite hierarchy of quasi-conservation laws. Whereas, the Hamiltonian deformation method is naturally embedded within this anomalous Lax formalism, where the deformation acts directly on the Hamiltonian functional rather than on the model's potential \cite{abhinav2}.  

In the next sections we provide the constructions of the deformations of the sine-Gordon, NLS and KdV models in the Riccati-type pseudopotential approach, respectively.

\section{Deformation of the sine-Gordon model}
\label{sec:dsg}

The compatibility condition  of 
a deformed system of Riccati-type Equations (\ref{compati1})-(\ref{AB11}), for specific coefficients $A_j$ and $B_j$ and relevant auxiliary fields $r$ and $s$, reproduces the equation of motion of a deformed sine-Gordon model (DSG). A DSG model has recently been considered as quasi-integrable in the anomalous zero-curvature approach \cite{jhep1}. We will present a dual Riccati-type formulation and construct explicitly the first two exact conservation laws and the first three  quasi-conservation laws. Moreover, we provide a pair of linear systems of equations for the DSG model and the associated inﬁnite tower of non-local conservation laws.  

Consider Lorentz-invariant field theories in $(1+1)$-dimensions with equations of motion expressed in light-cone coordinates $(\eta,\xi)$, given by \cite{jhep1}\footnote{In the $x$ and $t$ laboratory coordinates: $\eta=\frac{t+x}{2},\,\xi=\frac{t-x}{2},\, \pa_{\eta} = \pa_{t}+\pa_{x},\, \pa_{\xi} = \pa_{t}-\pa_{x},\, \pa_{\eta}\pa_{\xi}= \pa_{t}^2-\pa_{x}^2$.}
\br
\label{sg1}
\pa_\xi \pa_\eta \, w +  V^{(1)}(w) &= &0. 
\er
Here, $w$ is a real scalar field, $V(w)$ represents the scalar potential,  $\pa_{\xi}\,$ and $\pa_{\eta}$ denote partial derivatives, and $V^{(1)}(w) \equiv \frac{d}{dw} V(w)$. The family of potentials  $V(w)$ corresponds to specific deformations of the conventional sine-Gordon (SG) model, and Equation (\ref{sg1}) defines the deformed sine-Gordon (DSG) model equation of motion. The purpose is to explore the properties of this theory by employing modified techniques from integrable field theory, particularly through deformations of Riccati-type equations \cite{nucci1, nucci2}.

\subsection{Riccati-type pseudo-potential and conservation laws}

\label{sec:riccati1}

Next, let us consider the Riccati-type System (\ref{riccati11})-(\ref{riccati22}) with the following quantities \cite{npb1}
\br
\label{coef1}
A_0 &=& w_{\xi},\,\,\,\, A_1 = -2 \lambda^{-1},\,\,\,\,A_2 = w_{\xi},\\
B_0 &=& -\frac{1}{2} \lambda \frac{d}{dw} V(w)+ r,\,\,\,\,B_1 = - 2 \lambda (V(w) -2) -s,\,\,\,\,B_2 = \frac{1}{2} \lambda \frac{d}{dw} V(w).\label{coef2} 
\er
Notice that $\lambda$ is the spectral parameter and $r$ and $s$ are the auxiliary fields. Then, System (\ref{riccati11})-(\ref{riccati22}) becomes\cite{npb1, intech}
\br
\label{ricc1sg}
\pa_\xi u &=& -2 \lambda^{-1} \, u +\, \pa_\xi w + \, \pa_\xi w \,\, u^2,\\
\label{ricc2sg}
\pa_\eta u &=&-  2 \lambda\,  (V-2)\, u - \frac{1}{2} \,\lambda\, V^{(1)} + \frac{1}{2} \lambda\, V^{(1)}\, u^2 + r - u s.
\er
With the coefficients (\ref{coef1})-(\ref{coef2}), Equation (\ref{AB11}) can be written as
\br
\Delta + 
\Big[\lambda X -2 \lambda^{-1} r + (s + u\, r) \pa_\xi w - \pa_\xi r\Big] +
  \Big[\pa_\xi s + (r - u \, s )\pa_\xi w -\lambda u X \Big] \, u+ 
2 \Delta\, u^2 = 0, \label{poly12}
\er
with 
\br
X & \equiv & \pa_{\xi} w \( \frac{V^{(2)}}{2}+2 V -4\),\,\,\,\,\,\,V^{(2)} \equiv \frac{d^2}{d w^2} V(w),  \label{Xanomsg}\\
\Delta &\equiv& \pa_\xi \pa_\eta \, w +  V^{(1)}(w). \label{DSG11}
\er
Next, taking into account the following equations for the auxiliary fields $r(\xi,\eta)$ and  $s(\xi,\eta)$
\br
\label{ricc1r}
\pa_\xi r &=& -2 \lambda^{-1} \, r +\, \pa_\xi w  (u \, r + s)+ \lambda X,\\
\label{ricc2s}
\pa_\xi s &=& \pa_\xi w  (u\, s - r)+ \lambda X u,
\er  
into Equation (\ref{poly12}) one notices that the terms inside the brackets $\big[...\big]$ identically vanish. So, in (\ref{poly12}) we are left with a second order polynomial equation in $u$ of the form $\Delta + 2 \Delta\,  u^2$, which must vanish identically, i.e. $\Delta =0$. Then, one gets the equation of motion of the modified SG model (\ref{sg1}).
 
So, one has a set of two deformed Riccati-type equations for the pseudo-potential $u$ (\ref{ricc1sg})-(\ref{ricc2sg}) and System (\ref{ricc1r})-(\ref{ricc2s}) for the auxiliary fields $r$ and $s$. 

Notice that, for the usual SG model one has the potential 
\br
\label{usg}
V = -2 \cos{(2 w)} + 2,
\er
such that $X=0$, and so the auxiliary System (\ref{ricc1r})–(\ref{ricc2s}) admits the trivial solution  $r=s=0$. Taking into account this solution and considering the potential (\ref{usg}) into System (\ref{ricc1sg})–(\ref{ricc2sg}) one can get a pair of Riccati equations for the standard sine-Gordon (SG) model. These equations play a crucial role in analyzing the integrable properties of the SG model, including the derivation of an infinite set of conserved charges and B\"acklund transformations, which relate the field $w$ to another solution $\bar{w}$ \cite{prl1, wadati1}.
 
Note that only the $\eta-$component $\pa_{\eta} u$ of the Riccati equation associated with the ordinary sine-Gordon equation has been modified, deviating from the SG potential (\ref{usg}). This component encapsulates all information regarding the deformation of the model, which is encoded in the potential $V(w)$  and the auxiliary fields $r(\xi,\eta)$ and  $s(\xi,\eta)$. In contrast, the $\xi-$component $\pa_{\xi}u $, retains the same form as in the standard Riccati equation of the SG model.

As a verification one can compute the compatibility condition $[\pa_{\eta}\pa_\xi  u - \pa_\xi \pa_{\eta} u ]=0$ for the Riccati-type Equations (\ref{ricc1sg})-(\ref{ricc2sg}), and taking into account  the auxiliary System (\ref{ricc1r})-(\ref{ricc2s}) one can  rederive  the equation of motion of the deformed sine-Gordon model (\ref{sg1}).
 
It is important to highlight that, in the standard sine-Gordon (SG) model, System  (\ref{ricc1r})-(\ref{ricc2s}) admits the trivial solution $X=0 \rightarrow r=s=0$. Moreover, the existence of the Lax pair in the conventional SG model is reflected in its equivalent Riccati-type representation, given by System (\ref{ricc1sg})-(\ref{ricc2sg}) with the well known potential (\ref{usg}) \cite{nucci1, nucci2, prl1, wadati1}.

Next, we examine the relevant conservation laws within the framework of the Riccati-type System (\ref{ricc1sg})-(\ref{ricc2sg})  and the auxiliary Equations (\ref{ricc1r})-(\ref{ricc2s}). By substituting the $u^2$ from (\ref{ricc1sg}) into (\ref{ricc2sg})  and taking into account (\ref{ricc2s}), we obtain the following relationship 
\br
\label{qcons1}
\pa_{\eta} \(u \, \pa_{\xi} w \) + \pa_{\xi} \Big[ \lambda \, (V-2) - \frac{1}{2}\lambda \, u \, V^{(1)} \Big] = -   \pa_{\xi}  s  . 
\er

This identity can be regarded as a quasi-conservation law, and in fact, multiple such expressions can be formulated. Here, we construct one in which the non-homogeneous term on the right-hand side explicitly involves the deformation variable $s$. This choice ensures that, by setting $s=0$, the left-hand side recovers, order by order in $\lambda$, the polynomial conservation laws of the standard sine-Gordon (SG) model. Naturally, it is also possible to write an alternative equation that incorporates both deformation variables $s$ and $r$. 

Equation (\ref{qcons1}) will be used to reveal an infinite tower of conservation laws for the modified SG model (\ref{sg1}). However, its genuine conservation law nature remains to be confirmed. If the function $s(\xi, \eta)$ on the right-hand side of (\ref{qcons1}) contained non-local terms like$\int d\xi (...)$, the conservation property would break down, leading to anomalies similar to those in anomalous zero-curvature formulations of deformed Lax pairs and quasi-conservation laws \cite{jhep1}. Below, we demonstrate that the right-hand side of (\ref{qcons1}) can be expressed as $ [-\pa_{\xi}  s ] \equiv \pa_{\xi} S + \pa_{\eta} R$, where $S$ and $R$ are local functions of $w$ and its $\xi$ and  $\eta-$derivatives. Thus, a local expression for  $\pa_{\xi} s$ exists, ensuring that (\ref{qcons1}) defines a proper local conservation law.

Next, let us consider the expansions
\br
\label{expansg}
 u = \sum_{n=1}^{\infty} u_n \, \lambda^n,\,\,\,\,  s= \sum_{n=0}^{\infty} s_n \, \lambda^{n+2},\,\,\,\,  r= \sum_{n=0}^{\infty} r_n \, \lambda^{n+2}.
\er
Observe that the lowest order terms in $\lambda$ in the above expansions are chosen differently to ensure the consistency of the systems of equations (\ref{ricc1sg})–(\ref{ricc2sg}) and (\ref{ricc1r})–(\ref{ricc2s}). Specifically, the powers in the spectral parameter $\lambda$ must be compatible with the corresponding equations at each order in $\lambda^n$. Note that in (\ref{ricc1sg}) the initial terms corresponds to $\lambda^0$, providing $u_1 = \frac{1}{2} \pa_{\xi} w$ (see the appendix A in \cite{npb1}).  

The coefficients $u_n$ of the expansion above  can be determined order by order in powers of $\lambda$ from  the Riccati Equation (\ref{ricc1sg}). We refer to the  appendices in \cite{npb1} for the recursion relation for the $u_n\, 's$ and the expressions for the first $u_n$. Likewise, using the results for the $u_n\, 's$ one can get the relevant expressions for the $r_n\, 's$ and $s_n\, 's$ from (\ref{ricc1r})-(\ref{ricc2s}). 

So, replacing  those expansions into  the eq. (\ref{qcons1}) one has that the coefficient of the $n'$th order term becomes
\br
\label{anoconssg}
\pa_{\eta} a_{\xi}^{(n)} &+& \pa_{\xi} a_{\eta}^{(n)} = -   \pa_{\xi} s_{n-2},\,\,\,\,\,n=1,2,3,....;\, s_{-1}\equiv 0,\, \pa_{\xi} s_{0} =0,\\
a_{\xi}^{(n)} &\equiv & u_n \pa_{\xi} w,\,\,\,\,a_{\eta}^{(n)} \equiv   (V-2) \delta_{1, n}  - \frac{1}{2} u_{n-1} V^{(1)} ,\,\,\,\,\,u_0 \equiv 0.
\er 
Let us emphasize that only the coefficients $s_n$ appear on the right-hand side of Equation (\ref{anoconssg}), as the deformation variable $r_n$ does not enter the quasi-conservation Equation (\ref{qcons1}).

So, from Equation (\ref{anoconssg}), one has that order by order in powers of $\lambda$, the first two exact conservations laws ($n=1,2$) and the first three quasi-conservation laws ($n=3,4,5$) become  
\br
\label{n11}
n&=&1,\,\,\,\,\,\,\,\,\,\,\,\pa_{\eta} \( \frac{1}{2} (\pa_{\xi} w)^2\) + \pa_{\xi} \( V-2 \) =0,\\
\label{n21}
n&=&2,\,\,\,\,\,\,\,\,\,\,\,\,\frac{1}{4}\pa_{\xi} \Big[\pa_{\eta} \( \frac{1}{2} (\pa_{\xi} w)^2\) + \pa_{\xi} \( V-2 \)\Big] = 0, \\
\nonumber
n&=&3,\,\,\,\,\,\,\,\,\,\,\,\pa_{\eta} \Big[ \frac{1}{8}\pa_{\xi} w \( (\pa_{\xi} w)^3 + \pa_{\xi}^3 w \)\Big] + \pa_{\xi} \( \frac{1}{8}  \pa_{\xi}^2 w V^{(1)} \) =  -\pa_{\xi} s_1, \\
\label{n31}
&&\\
 \nonumber
n&=&4,\,\,\,\,\,\,\,\,\,\,\,\,\,
\pa_{\eta} \( \frac{5}{16} \pa_{\xi}^2 w (\pa_{\xi} w)^3 + \frac{1}{16}  \pa_{\xi} w \pa_{\xi}^4 w \) + \pa_{\xi} \( \frac{1}{16} V^{(1)} [ (\pa_{\xi} w)^3 + \pa_{\xi}^3 w] \) = - \pa_{\xi} s_{2},\\
\label{n41}
&&\\
\nonumber
n&=&5,\,\,\,\,\,\,\,\,\,
\pa_{\eta} \( \frac{11}{32}(\pa_{\xi}^2 w)^2 (\pa_{\xi} w)^2 + \frac{7}{32}  \pa_{\xi}^3 w (\pa_{\xi} w)^3  +  \frac{1}{16}(\pa_{\xi} w)^6 + \frac{1}{32}\pa_{\xi} w \pa_{\xi}^5 w \)  -  \pa_{\xi} \Big[ \frac{1}{2} V^{(1)} u_4 \Big] = -\pa_{\xi} s_3,\\
\label{n51}
&& 
\er
with 
\br
-\pa_{\xi} s_1 &=& \frac{1}{4} X  \pa_{\xi}^2 w  -\frac{1}{4}  \pa_{\xi} w \, \pa_{\xi} X,
\label{n3r}\\
-\pa_{\xi} s_{2} &=&  \frac{1}{8} X \pa_{\xi}^3 w - \frac{1}{8}  \pa_{\xi} w \pa_{\xi}^2 X, \label {n4r}
\\
 -\pa_{\xi} s_3&=&-\frac{3}{16} (\pa_{\xi} w)^3 \pa_{\xi} X+ \frac{1}{16}
\( \pa_{\xi}^4 w + \pa_{\xi}[(\pa_{\xi} w)^3] \) \,  X - \frac{1}{16} \pa_{\xi} w \pa_{\xi}^3 X. \label {n5rr}
\er

Notice that, the right hand sides in (\ref{n3r})-(\ref{n5rr}), which also appear in the r.h.s. of (\ref{n31})-(\ref{n51}), can be rewritten as  
\br
\label{s1a}
-\pa_{\xi} s_1  
 &=& \pa_{\eta} R_1 + \pa_{\xi} S_1, \label{s1b}
  \\
\label{s1c}
R_1 &\equiv &-\frac{1}{8} [\pa_{\xi}^2 w]^2 + \frac{1}{8} [\pa_{\xi} w]^4 ,\,\,S_1 \equiv - \frac{1}{8} [\pa_{\xi} w]^2 V^{(2)},\label{rs22}\\
\label{s2a}
-\pa_{\xi} s_{2} &=& \frac{1}{8} \pa_{\xi} [\pa_\xi^2 w X - \pa_\xi w \pa_\xi X],\\
-\pa_{\xi} s_3 &=& \pa_{\eta} R_3 + \pa_{\xi} S_3, \label{s3a}
\er
where the explicit expressions of $R_3$ and $S_3$ are provided in the eqs. (E.1)-(E.3) of the appendix E of \cite{npb1}. Notice that the right hand sides in (\ref{s1a})-(\ref{s3a}) are written as $(\pa_{\eta}R_k+\pa_{\xi} S_k),\, (k=1,2,3)$. 

Next, let us discuss the equations above. 

{\bf First order $n=1$}. Notice that the r.h.s. of (\ref{anoconssg}) vanishes at this order, i.e. by definition one has $ s_{-1}\equiv 0$. In fact, the conservation law (\ref{n11}) provides the first conserved charge
\br
\label{nnq1}
q^{(1)} &=& \int dx\,\Big[ \frac{1}{2} ( \pa_{\xi} w)^2 + (V-2) \Big].
\er
Equations (\ref{n11}) and (\ref{nnq1}), together with their duals and the relevant charge $ \widetilde{q}^{(1)} = \int dx\, [ \frac{1}{2} ( \pa_{\eta} w)^2 + (V-2) ]$ which will be presented below, give rise to the energy and momentum charges written in laboratory coordinates $(x,t)$, respectively, as
\br
\label{energy1}
E &=& \frac{1}{2} \Big[ q^{(1)} + \widetilde{q}^{(1)}\Big],\\
  &=& \int dx \Big[ \frac{1}{2} (\pa_x w)^2 + \frac{1}{2} (\pa_t w)^2 + (V-2) \Big],
  \label{energy11}\\
\label{momentum1}
P &=&  \frac{1}{2} \Big[\widetilde{q}^{(1)}-  q^{(1)} \Big],\\
  &=&\int dx [ \pa_x w\pa_t w ] .\label{momentum11} 
\er

{\bf Second order $n=2$}.

At this order ${\cal O}(\lambda^{2})$ one can write
\br
\label{n2sg}
\frac{1}{4}\pa_{\xi} \Big[\pa_{\eta} \( \frac{1}{2} (\pa_{\xi} w)^2\) + \pa_{\xi} \( V-2 \)\Big] = 0. 
\er
As usual, we can define the charge 
\br
\label{nnq2}
q^{(2)} &=& \int dx\, \frac{1}{4} \(\pa_{\xi}^2 w \pa_{\xi} w +  \pa_{\xi} w V^{(1)}\),\\
&=&\frac{1}{4} \frac{d}{dt} \( E - P\).\label{nnq21}\er 
So, the eq. (\ref{n2sg}) does not provide an independent new charge in laboratory coordinates $(x,t)$. So, there is no an independent new charge at this order. Notice that also the usual SG model does not possess an independent charge at this order \cite{sanuki}. 
From this point forward and for the higher order charges the term encoding the deformation away from the usual SG model, i.e. the r.h.s. of (\ref{anoconssg}), will play an important role in the construction of the conservation laws.

{\bf Third order $n=3$}.

Taking into account (\ref{n31}) and (\ref{s1a}) the conservation law turns out to be  
\br
\label{nn3}
\frac{1}{8}\pa^{2}_{\xi}\Big[ \pa_{\eta} \( \frac{1}{2} (\pa_{\xi} w)^2\) + \pa_{\xi} \( V-2 \) \Big]=0.
\er
Notice that this form of the third order conservation law holds strictly for deformed SG models, i.e. for models such that $X \neq 0$. In the usual SG model the eq. (\ref{n31}) with vanishing r.h.s., since in that case $X\equiv 0$ implies $\pa_{\xi}s_1 =0$ in (\ref{n3r}), provides the relevant conservation law at this order. 

Next, the charge which follows from the above conservation law (\ref{nn3}) becomes
\br
q^{(3)} = \frac{1}{8} \frac{d^2}{dt^2} \( E - P\).\label{ch32} \er 

Thus, at this order and within this formulation, unlike the conventional SG model, the deformed SG model (\ref{sg1}) does not possess an independent conserved charge.

Nevertheless, it can be shown that the third-order charge and anomaly defined in \cite{jhep1, jhep2, cnsns} can be reformulated in our notation as follows
\br
\label{qa3}
q_{a}^{(3)} &=& \int dx [ \frac{1}{8}(\pa_{\xi} w)^4 +\frac{1}{8} \pa_{\xi} w\pa_{\xi}^3 w+  \frac{1}{8}  \pa_{\xi}^2 w V^{(1)} + \frac{1}{4} X \pa_{\xi}w  ], \er
  and 
\br
\label{betaxieta}
\beta^{(3)} \equiv \frac{1}{2} \pa_{\xi}^2w  X = \pa_{\xi} S_1 + \pa_{\eta} R_1 + \frac{1}{4} \pa_{\xi} \( \pa_{\xi}w\, X\),\er
with $R_1, S_1$ given in (\ref{rs22}). So, one has 
\br
\label{qa3ano}
\frac{d}{dt} q_{a}^{(3)} = \int_{-\infty}^{+\infty} dx \beta^{(3)}.
\er

The charge $q_{a}^{(3)}$ defined in (\ref{qa3}) has been previously computed through numerical simulations of two-soliton collisions for a specific deformation of the SG model \cite{npb1}. Interestingly, the numerical simulations demonstrate that the charge $q_{a}^{(3)}$ are exactly conserved in general two-soliton configurations, with no observable deviation within the precision of the numerical method.

{\bf Fourth order $n=4$}
   
Taking into account (\ref{n41}) and (\ref{s2a}) one has the conservation law
\br
\label{n4ii}
\pa_{\eta} \( \frac{5}{16} \pa_{\xi}^2 w (\pa_{\xi} w)^3 + \frac{1}{16}  \pa_{\xi} w \pa_{\xi}^4 w \) + \pa_{\xi} \( \frac{1}{16}  [ (\pa_{\xi} w)^3 + \pa_{\xi}^3 w]V^{(1)} - \frac{1}{8}  [\pa_\xi^2 w X - \pa_\xi w \pa_\xi X]   \) &=& 0.   
\er
 
Next, from (\ref{n4ii}) one can define the charge 
\br
q^{(4)} &=& \int dx \,\Big\{\frac{5}{16} \pa_{\xi}^2 w (\pa_{\xi} w)^3 + \frac{1}{16}  \pa_{\xi} w \pa_{\xi}^4 w  +   \frac{1}{16}  [ (\pa_{\xi} w)^3 + \pa_{\xi}^3 w]V^{(1)} - \frac{1}{8}  [\pa_\xi^2 w X - \pa_\xi w \pa_\xi X]  \Big\}, \\
& \equiv & 0 \label{q41}.
\er
Thus, it has been demonstrated that this charge identically vanishes under appropriate boundary conditions.

{\bf Fifth order $n=5$}

Similarly, by considering (\ref{n51}) and (\ref{s3a}) and performing a lengthy calculation, the fifth-order conservation law becomes
\br
 \frac{1}{32}\pa_{\xi}^4 \Big\{ \frac{1}{2} \pa_{\eta} (\pa_{\xi} w)^2  + \pa_{\xi} (V-2) \Big\} = 0 \label{cons5}.
\er 
From the above conservation law it follows the fifth order  conserved charge
\br
\label{q5}
q^{(5)} & \equiv &\frac{1}{32} \frac{d^4}{dt^4}  \int dx\,\Big[ \frac{1}{2} ( \pa_{\xi} w)^2 + (V-2) \Big] , \\
 &\equiv& \frac{1}{32}  \frac{d^4}{dt^4}  \( E - P \)\label{q51}.
\er 

Thus, the fifth-order charge $q^{(5)}$ in (\ref{q5}) is not an independent charge of the deformed sine-Gordon model (\ref{sg1}), despite arising from a genuine conservation law in the Riccati-type formulation, beyond energy and momentum.

Below, we define a related embedded charge $q^{(5)}_a$ along with its corresponding anomaly term $\beta^{(5)}$, i.e.
\br
\frac{d}{dt} q_{a}^{(5)} = \int_{-\infty}^{+\infty} dx \beta^{(5)}.
\er 
The  `anomaly' term $\beta^{(5)}$ introduced in \cite{jhep1} has been  written, in the notation of \cite{npb1}, as a  term of the r.h.s. of (\ref{n5rr}) as 
\br
\label{anomaly50}
 -2^4 \pa_{\xi} s_3&=& 2\beta^{(5)} - \pa_{\xi}[3 (\pa_{\xi} w)^3 X + \pa^3_{\xi} w X -  \pa_{\xi}^2 w \pa_{\xi} X +  \pa_{\xi} w \pa^2_{\xi} X ], \\
\beta^{(5)} & \equiv &  \frac{1}{16} [ \pa_{\xi}^4 w + 6 (\pa_{\xi} w)^2 \pa_{\xi}^2 w ] \,  X . \label{anomaly500}
\er 

Therefore, the additional terms $\pa_{\xi} [.\,.\,.]$ appearing in the expression of $(- 2^4 \pa_{\xi} s_3)$ and provided in (\ref{anomaly50}) can be incorporated into the l.h.s. of the  conservation law (\ref{n51}). Since the anomaly $\beta^{(5)}$ can be written in the form $\pa_{\eta} [.\,.\,.] + \pa_{\xi} [.\,.\,.] $, one can define the asymptotically conserved charge $q^{(5)}_a$ as
\br
\frac{d}{dt}q^{(5)}_a  &=& 2 \int  dx\,  \beta^{(5)} , \\
 &=& \frac{1}{8} \frac{d}{dt} \int  dx \, \Big\{  \frac{1}{2}  (\pa_{\xi} w)^6 + \frac{1}{4}  (\pa^3_{\xi} w)^2- \frac{5}{2}  (\pa^2_{\xi} w)^2 (\pa_{\xi} w)^2 - 6  (\pa_{\xi} w)^4 + 6 (\pa^2_{\xi} w)^2 - \frac{1}{2}  \pa^4_{\xi} w  V^{(1)} + \nonumber \\ \nonumber
&& 2 [ 2  \pa_{\xi} w \pa^3_{\xi} w  + \frac{1}{2} (\pa^2_{\xi} w)^2 +  \frac{3}{2} (\pa_{\xi} w)^4 ] V \Big\} + \\
&&  \frac{1}{8} \frac{d^2}{dt^2} \int  dx\,  \Big\{\frac{1}{2}  \pa^3_{\xi} w V^{(1)}  - 2 \pa_{\xi} w\pa^2_{\xi} w V \Big\} - \frac{1}{2}  \frac{d^3}{dt^3}  \int dx \,  \frac{1}{2} ( \pa_{\xi} w)^2. \label{charge5sg}
\er
 Finally, the fifth-order quasi-conservation law from \cite{jhep1} can be expressed as an exact conservation law, provided that the form (\ref{s3a}) is utilized. This manipulation results in the exact conservation law (\ref{cons5}).

The charge $q_{a}^{(5)}$ has been previously computed through numerical simulations of two-soliton collisions for a specific deformation of the SG model \cite{npb1}. These charges can be considered, in general, as asymptotically conserved ones.
 
In this way the r.h.s.'s  $[-\pa_{\xi} s_{n} \, (n=1,2,3,4)]$ of the relevant conservation laws have  been written  as $\pa_{\eta} R_n + \pa_{\xi} S_n$. It has been shown that this property holds in general for each term  $[(-\pa_{\xi} s_{n}),\,n\geq 1$], of the tower of conservation laws. So, the conservation laws  (\ref{anoconssg}) in general can be written as   
 \br
\label{anocons1}
\pa_{\eta} [ a_{\xi}^{(n)} - R_{n-2} ] + \pa_{\xi} [ a_{\eta}^{(n)} - S_{n-2}] = 0,\,\,\,\,\,n=1,2,3,...( S_{k}=R_{k}\equiv 0,\,\,\,k=-1,0).
\er
  
Therefore, the construction provides an infinite tower of conservation laws (\ref{anocons1}). 

The notion of quasi-integrability and its associated asymptotically conserved charges, originally introduced in \cite{jhep1} and further developed in \cite{cnsns, jhep4, proc1, jhep5} through the introduction of subsets of exactly conserved charges, depends on the specific field configurations to which it is applied—such as kink-kink, kink-antikink, and breather configurations in the deformed model. This contrasts with the conventional concept of integrability, where conserved charges are defined for all fields of the model.

\subsection{Riccati-type pseudo-potential and dual conservation laws}

\label{sec:riccatidual}

Since the deformed SG model (\ref{sg1}) remains invariant under the transformation $\eta \leftrightarrow \xi$, there naturally exists a dual Riccati-type formulation corresponding to System  (\ref{ricc1sg})-(\ref{ricc2s}) discussed above. Accordingly, it is introduced the following system of equations for the new pseudo-potential $\widetilde{u}$ \cite{npb1}
\br
\label{ricc11d}
\pa_\eta \widetilde{u} &=& -2 \lambda \, \widetilde{u} +\, \pa_\eta w + \, \pa_\eta w \,\, \widetilde{u}^2,\\
\label{ricc22d}
\pa_\xi \widetilde{u} &=&-  \frac{2}{\lambda}\,  (V-2)\, \widetilde{u} - \frac{1}{2 \lambda}  \, V' + \frac{1}{2 \lambda} \, V'\, \widetilde{u}^2 + \widetilde{r} - \widetilde{u} \, \widetilde{s}. 
\er
Observe that the transformations $\lambda \rightarrow  \lambda^{-1}$ and $\xi \leftrightarrow \eta$ have been performed into System (\ref{ricc1sg})-(\ref{ricc2sg}), while also relabeling the pseudo-potential and auxiliary fields. Meanwhile, the field $w$  and the deformed sine-Gordon potential $V(w)$ remain unchanged.

Next, the following equations for the auxiliary fields $\widetilde{r}(\xi,\eta)$ and  $\widetilde{s}(\xi,\eta)$ are considered
\br
\label{ricc1rd}
\pa_\eta \widetilde{r} &=& -2 \lambda \, \widetilde{r} +\, \pa_\eta w  (\widetilde{u} \,\widetilde{r} + \widetilde{s})+ \lambda^{-1} \widetilde{X},\\
\label{ricc2sd}
\pa_\eta \widetilde{s} &=& \pa_\eta w  (\widetilde{u} \, \widetilde{s} - \widetilde{r})+ \lambda^{-1} \widetilde{X} \widetilde{u},\\
\widetilde{X} & \equiv & \pa_{\eta} w \( \frac{V^{(2)}}{2}+2 V -4\),\,\,\,\,\,\,V^{(2)} \equiv \frac{d^2}{d w^2} V(w).  \label{Xanomd}
\er  
Thus, one has two deformed Riccati-type equations for the pseudo-potential $\widetilde{u}$ (\ref{ricc11d})-(\ref{ricc22d}), along with System (\ref{ricc1rd})-(\ref{ricc2sd}) governing the auxiliary fields $\widetilde{r}$ and  $\widetilde{s}$. Moreover, for the specific potential (\ref{usg}), $\widetilde{X}$
vanishes identically, and the linear System (\ref{ricc11d})-(\ref{ricc22d}) reduces to the ordinary sine-Gordon integrable model, provided that $\widetilde{s}=\widetilde{r}=0$ in (\ref{ricc22d}).

Similarly to the previous subsection, substituting the expression for $\widetilde{u}^2$ from (\ref{ricc11d}) into (\ref{ricc22d}) yields the following relationship
\br
\label{qconsdsg}
\pa_{\xi} \(\widetilde{u} \, \pa_{\eta} w \) + \pa_{\eta} \( \lambda^{-1} \, (V-2) - \frac{1}{2}\lambda^{-1} \, \widetilde{u} \, V^{(1)} \) = - \pa_{\eta} \widetilde{s}. 
\er
 
This equation serves as a tool for revealing an infinite number of new conservation laws associated with the modified SG model (\ref{sg1}). Thus, it is considered the following expansions
\br
\label{expan2}
\widetilde{u} = \sum_{n=1}^{\infty} \widetilde{u}_n \, \lambda^{-n},\,\,\,\, \widetilde{s}= \sum_{n=0}^{\infty} \widetilde{s}_n \, \lambda^{-(n+2)},\,\,\,\,  \widetilde{r}= \sum_{n=0}^{\infty} \widetilde{r}_n \, \lambda^{-(n+2)}.
\er
The components $\widetilde{u}_n$
 are determined recursively by substituting the above expression into (\ref{ricc11d}), while the components $\widetilde{s}_n$ and $\widetilde{r}_n$ are obtained from System (\ref{ricc1rd})-(\ref{ricc2sd}). In the Appendices C and D of \cite{npb1} it is provided the explicit expressions for the first few $\widetilde{u}_n,\,\widetilde{s}_n, \widetilde{r}_n$.  Using these components, one can systematically derive the conservation laws order by order in powers of $\lambda^{-1}$ by substituting the expansions into (\ref{qconsdsg}). Consequently, the conservation law at order $(-n)$ takes the form
\br
\label{anocons2}
\pa_{\xi} \widetilde{a}_{\eta}^{(n)} &+& \pa_{\eta} \widetilde{a}_{\xi}^{(n)} = -   \pa_{\eta} \widetilde{s}_{n-2},\,\,\,\,\,n=1,2,3,....;\, \widetilde{s}_{-1}\equiv 0,\\
\widetilde{a}_{\eta}^{(n)} &\equiv & \widetilde{u}_n \pa_{\eta} w,\,\,\,\,\widetilde{a}_{\xi}^{(n)} \equiv   (V-2) \delta_{1, n}  - \frac{1}{2} \widetilde{u}_{n-1} V^{(1)} ,\,\,\,\,\,\widetilde{u}_0 \equiv 0. 
\er 
Next, we provide the first order conservation law 
\br
\label{n12}
\pa_{\xi} \( \frac{1}{2} (\pa_{\eta} w)^2\) + \pa_{\eta} \( V -2\) =0.
\er 
This equation furnishes the conserved charge
\br
\label{first1d}
\widetilde{q}^{(1)} = \int dx\, [ \frac{1}{2} (\pa_{\eta} w)^2 + (V-2)].
\er
This charge combined to its dual has been used to write the energy and momentum charges as in eqs. (\ref{energy1})-(\ref{momentum11}) in the last subsection.

The next order term becomes
\br
\label{n2d}
\pa_{\xi} \( \frac{1}{4} \pa_{\eta}^2 w \pa_{\eta} w\) + \pa_{\eta} \( \frac{1}{4} \pa_{\eta} w  V^{(1)}  \) = 0. 
\er
Notice that the r.h.s. vanishes due to $\pa_{\eta} \widetilde{s}_{0}=0$. As usual, it is defined the charge 
\br
\label{nn2qd}
\widetilde{q}^{(2)} &=& \int dx\, \frac{1}{4} \(\pa_{\eta}^2 w \pa_{\eta} w +  \pa_{\eta} w V^{(1)}\),\\
&=&\frac{1}{4} \frac{d}{dt} \( E + P\).\label{nnq2d1} \er 
Thus, Equation (\ref{n2d}) does not yield an independent new charge. Similarly to the dual case discussed in the previous subsection, no independent new charge arises at this order.

As in the construction of preceding subsection, from this point forward and for the higher order charges the terms encoding the deformation away from the usual SG model, i.e. $(-\pa_{\eta} \widetilde{s}_{n-2})$ in the r.h.s. of  (\ref{anocons2}), will play an important role. 

As in the construction outlined in the previous subsection, from this point onward and for higher-order charges, the terms representing the deformation from the usual SG model-specifically,$(-\pa_{\eta} \widetilde{s}_{n-2})$ in the r.h.s. of  (\ref{anocons2})-, will play a crucial role.

The next order term provides
\br
\label{n32}
\pa_{\xi} \Big[ \frac{1}{8}\pa_{\eta} w \( (\pa_{\eta} w)^3 + \pa_{\eta}^3 w \)\Big] + \pa_{\eta} \( \frac{1}{8}  \pa_{\eta}^2 w V^{(1)} \) =  -\pa_{\eta} \widetilde{s}_1.
\er 
The r.h.s. of (\ref{n32}) can be written as
\br
\label{s1ad}
-\pa_{\eta} \widetilde{s}_1 &=& \frac{1}{4} \widetilde{X}  \pa_{\eta}^2 w  -\frac{1}{4}  \pa_{\eta} w \, \pa_{\eta} \widetilde{X},\\
 &=& \pa_{\xi} \widetilde{R}_1 + \pa_{\eta} \widetilde{S}_1, \label{s1bd}
  \\
\label{s1cd}
\widetilde{R}_1 &\equiv &-\frac{1}{8} [\pa_{\eta}^2 w]^2 + \frac{1}{8} [\pa_{\eta} w]^4 ,\,\,\widetilde{S}_1 \equiv - \frac{1}{8} [\pa_{\eta} w]^2 V^{(2)}.\label{rs22d}
\er
Then, the conservation law (\ref{n32}) becomes  
\br
\label{nn3d}
\pa_{\xi} \Big[  \frac{1}{8}  \pa_{\eta} w \pa_{\eta}^3 w + \frac{1}{8} [\pa_{\eta}^2 w]^2   \Big] + \pa_{\eta} \Big[ \frac{1}{8}  \pa_{\eta}^2 w V^{(1)} +  \frac{1}{8} [\pa_{\eta} w]^2 V^{(2)}  \Big] = 0.
\er
Thus, the charge which follows from (\ref{nn3d}) becomes
\br
\label{ch31d}
\widetilde{q}^{(3)} &=& \int dx\, \Big[  \frac{1}{8}  \pa_{\eta} w \pa_{\eta}^3 w + \frac{1}{8} [\pa_{\eta}^2 w]^2   + \frac{1}{8}  \pa_{\eta}^2 w V^{(1)} +  \frac{1}{8} [\pa_{\eta} w]^2 V^{(2)}  \Big], \\
&=&\frac{1}{8} \frac{d^2}{dt^2} \( E + P\).\label{ch32d} \er 
So, in this dual formulation and at this order, unlike the ordinary SG model, the deformed SG model (\ref{sg1}) does not possess an independent conserved charge.

Moreover, as in (\ref{qa3ano}), it is shown that the third order charge and anomaly in \cite{jhep1, jhep2, cnsns}, in our notation, can be rewritten, respectively, in the form 
\br
\label{qa3anod}
\frac{d}{dt} \widetilde{q}_{a}^{(3)} = \int_{-\infty}^{+\infty} dx \widetilde{\beta}^{(3)}.
\er

\br
\widetilde{q}_{a}^{(3)} &=& \int dx [ \frac{1}{8}(\pa_{\eta} w)^4 +\frac{1}{8} \pa_{\eta} w\pa_{\eta}^3 w+  \frac{1}{8}  \pa_{\eta}^2 w V^{(1)} + \frac{1}{4} \widetilde{X} \pa_{\eta}w  ], \er
  and 
\br
\label{betaxietad}
\widetilde{\beta}^{(3)} \equiv \frac{1}{2} \pa_{\eta}^2 w  \widetilde{X} = \pa_{\eta} \widetilde{S}_1 + \pa_{\xi} \widetilde{R}_1 + \frac{1}{4} \pa_{\eta} \( \pa_{\eta} w\, \widetilde{X}\),\er
with $\widetilde{R}_1, \widetilde{S}_1$ given in (\ref{rs22d}). 

A relationship between the charge $\widetilde{q}_{a}^{(3)}$, its `anomaly'  $\widetilde{\beta}^{(3)}$ and the exactly conserved charge $\widetilde{q}^{(3)}$, follows from (\ref{n32}) or (\ref{nn3d})  
\br
\label{asy31d}
\widetilde{q}_{a}^{(3)} &=&  \widetilde{q}^{(3)} + \int \, dx  \Big[\widetilde{S}_1 + \widetilde{R}_1 + \frac{1}{4} \pa_{\eta}w \widetilde{X}\Big],\\
             &=&  \frac{1}{8} \frac{d^2}{dt^2} [ E + P] + \int \, dx  \Big[ \frac{1}{8} (\pa_{\eta} w)^4- \frac{1}{8} (\pa_{\eta}^2 w)^2 +  \frac{1}{2} (\pa_{\eta} w)^2[V -2] \Big]. \label{asy32d}
\er  
The charge $\widetilde{q}_{a}^{(3)}$, combined with its dual $q_{a}^{(3)}$ in (\ref{qa3}), has been computed in \cite{npb1} by numerical simulations of two-soliton collisions for a deformed SG model. 

The next order terms ${\cal O}(\lambda^{-4})$ and ${\cal O}(\lambda^{-5})$ become, respectively
\br
  \label{n4id}
\pa_{\xi} \( \frac{5}{16} \pa_{\eta}^2 w (\pa_{\eta} w)^3 + \frac{1}{16}  \pa_{\eta} w \pa_{\eta}^4 w \) + \pa_{\eta} \( \frac{1}{16} V^{(1)} [ (\pa_{\eta} w)^3 + \pa_{\eta}^3 w] \) = \pa_{\eta} \widetilde{s}_{2} .
\er
with  
\br
\pa_{\eta} \widetilde{s}_{2} &=&  \frac{1}{8} \widetilde{X} \pa_{\eta}^3 w - \frac{1}{8}  \pa_{\eta} w \pa_{\eta}^2 \widetilde{X},\\
 &=& \frac{1}{8} \pa_{\eta} [\pa_\eta^2 w \widetilde{X} - \pa_\eta w \pa_\eta \widetilde{X}]. \label{an4id}
\er
and 
\br
\label{n5d}
\pa_{\xi} \( \frac{11}{32}(\pa_{\eta}^2 w)^2 (\pa_{\eta} w)^2 + \frac{7}{32}  \pa_{\eta}^3 w (\pa_{\eta} w)^3  +  \frac{1}{16}(\pa_{\eta} w)^6 + \frac{1}{32}\pa_{\eta} w \pa_{\eta}^5 w \)  -  \pa_{\eta} \Big[ \frac{1}{2} V^{(1)} \widetilde{u}_4 \Big] = -\pa_{\eta} \widetilde{s}_3,\er
with
\br
 -\pa_{\eta} \widetilde{s}_3&=&-\frac{3}{16} (\pa_{\eta} w)^3 \pa_{\eta} \widetilde{X}+\frac{1}{16}
\( \pa_{\eta}^4 w + \pa_{\eta}[(\pa_{\eta} w)^3] \) \,  \widetilde{X}- \frac{1}{16} \pa_{\eta} w \pa^3_{\eta} \widetilde{X}\\
  &=& \pa_{\xi} \widetilde{R}_3 + \pa_{\eta} \widetilde{S}_3, 
 \er
where $\widetilde{R}_3$ and $\widetilde{S}_3$ are provided  in the appendix F of \cite{npb1}. 
 
Equation (\ref{n4id}), taking into account (\ref{an4id}), can be written as 
\br
\label{n4iid}
\pa_{\xi} \( \frac{5}{16} \pa_{\eta}^2 w (\pa_{\eta} w)^3 + \frac{1}{16}  \pa_{\eta} w \pa_{\eta}^4 w \) + \pa_{\eta} \( \frac{1}{16}  [ (\pa_{\eta} w)^3 + \pa_{\eta}^3 w]V^{(1)} - \frac{1}{8}  [\pa_\eta^2 w \widetilde{X} - \pa_\eta w \pa_\eta \widetilde{X}]   \) = 0.   
\er 
From (\ref{n4iid}) one can define the trivial charge 
\br
\widetilde{q}^{(4)} &=& \int dx \,\Big\{\frac{5}{16} \pa_{\eta}^2 w (\pa_{\eta} w)^3 + \frac{1}{16}  \pa_{\eta} w \pa_{\eta}^4 w  +   \frac{1}{16}  [ (\pa_{\eta} w)^3 + \pa_{\eta}^3 w]V^{(1)} - \frac{1}{8}  [\pa_\eta^2 w \widetilde{X} - \pa_\eta w \pa_\eta \widetilde{X}]  \Big\}, \\
& \equiv & 0. \label{q4d}
\er 
Thus, it has been demonstrated that the dual charge at this order also vanishes identically under suitable boundary conditions.
   
Next, a lengthy calculation enables us to express Equation (\ref{n5d}) as the fifth-order conservation law
\br
\frac{1}{32}\pa_{\eta}^4 \Big\{ \frac{1}{2} \pa_{\xi} (\pa_{\eta} w)^2  + \pa_{\eta} (V-2) \Big\} = 0 \label{cons5d}.
\er 
From the above conservation law it follows the fifth order  conserved charge
\br
\label{q5d}
\widetilde{q}^{(5)} & \equiv &\frac{1}{32} \frac{d^4}{dt^4}  \int dx\,\Big[ \frac{1}{2} ( \pa_{\eta} w)^2 + (V-2) \Big] , \\
 &\equiv& \frac{1}{32}  \frac{d^4}{dt^4}  \( E + P \). \label{q5d1}
\er 
Thus, the fifth-order dual charge $\widetilde{q}^{(5)}$ in (\ref{q5d}) is not an independent charge of the deformed sine-Gordon model (\ref{sg1}), despite arising from a genuine conservation law in the Riccati-type dual formulation, beyond energy and momentum.

In this way,  the r.h.s.'s  $[-\pa_{\eta} \widetilde{s}_{n} \, (n=1,2,3,4)]$ of the dual conservation laws have  been written as $\pa_{\xi} \widetilde{R}_n + \pa_{\eta} \widetilde{S}_n$. So, the conservation laws  (\ref{anocons2}) can be written as   
 \br
\label{anocons1d}
\pa_{\xi} [ \widetilde{a}_{\eta}^{(n)} - \widetilde{R}_{n-2} ] + \pa_{\eta} [ \widetilde{a}_{\xi}^{(n)} - \widetilde{S}_{n-2}] = 0,\,\,\,\,\,n=1,2,3,...( \widetilde{S}_{k}= \widetilde{R}_{k}\equiv 0,\,\,\,k=-1,0).
\er
Thus, beyond the energy-momentum charges, the higher-order asymptotically conserved charges retain the same form as those in the usual sine-Gordon model, despite being governed by the dynamics of the deformed sine-Gordon model with the potential $V(w)$, which supports solitary waves. 

The existence of an infinite number of conservation laws is one of the defining characteristics of integrable models, as they impose strong constraints on the system's dynamics and enable the presence of soliton-type solutions. As discussed earlier, in the context of deformed SG models, conservation laws can be derived directly from specific structures, such as the system’s deformed Riccati-type equations or the abelianization procedure in the anomalous Lax pair formulation \cite{jhep1}. However, rigorously proving the mutual independence and non-triviality of the charges associated with the conservation laws (\ref{anocons1}) is often a challenging task. In the Riccati-type pseudo-potential formulation, these charges do not automatically satisfy these criteria, requiring a detailed order-by-order examination of their non-triviality and mutual independence.

So, let us examine the first order charges. The first order dual conserved charges for $n=1$ in (\ref{nnq1}) and (\ref{first1d}), whose linear combinations provide the energy $E$ in (\ref{energy1}) and momentum $P$ in (\ref{momentum1}) are, in fact, non-trivial and mutually independent. 

The dual conserved charges for $n=2$ in (\ref{nnq21}) and (\ref{nnq2d1}), for $n=3$ in (\ref{ch32}) and (\ref{ch32d}), and for $n=5$ in (\ref{q51}) and (\ref{q5d1}), are not independent charges, since they appear to be $(n-1)'th$ order time derivatives of the form $\frac{1}{2^n} \frac{d^{n-1}}{dt^{n-1}}(E \pm P),\,n=2,3,5$. Moreover, for $n=4$ the both dual charges (\ref{q41}) and (\ref{q4d}) become vanishing trivial charges, respectively.  

However, associated to $n=3$ it is possible to define the asymptotically conserved charges $q^{(3)}_a$ and $\widetilde{q}^{(3)}_a$ as in (\ref{qa3}) and  (\ref{asy32d}), respectively. Similar expressions $q^{(5)}_a$ and $\widetilde{q}^{(5)}_a$ can be defined for $n=5$ (see Equations (4.48) and (4.93) of \cite{npb1}).

\subsection{Pseudo-potentials and a linear system associated to DSG}
\label{sec:linear}

In this subsection it is addressed the problem of formulating a linear system of equations associated with the DSG model. It is proposed the next linear system \cite{npb1}
\br
\pa_{\xi} \Phi &=& A_{\xi} \Phi, \label{lin1}\\
\pa_{\eta} \Phi &=& A_{\eta} \Phi, \label{lin2}\\
A_{\xi}&\equiv& a_0 + \lambda \, a_1,\,\,\,\,\,\,A_{\eta}\equiv b_0 + \lambda \, b_1,\label{AB}\\
a_0& \equiv & - 2 \frac{ (\pa_{\xi} w)^3}{\pa_{\xi}^2 w},\,\,\,\,b_0 \equiv \zeta = \int^{\xi} d\xi' \Big[6 V^{(1)}  \frac{(\pa_{\xi'} w)^2}{\pa_{\xi'}^2 w}
- 2 V^{(2)}   \frac{(\pa_{\xi'} w)^4 }{(\pa_{\xi'}^2 w)^2}\Big],\label{ab00}\\
a_1 &\equiv &  \frac{1}{2} (\pa_{\xi} w)^2,\,\,\,b_1 \equiv - 2 - V.  
\label{ab11}
\er
The compatibility condition of the linear problem (\ref{lin1})-(\ref{lin2}) provides  
\br
\label{zeroeq}
   \Delta(\xi,\eta)\, \lambda - 6 \frac{\pa_{\xi} w}{\pa_{\xi}^2 w}  \Delta(\xi,\eta) + 2 
\frac{(\pa_{\xi} w)^2}{(\pa_{\xi}^2 w)^2} \pa_{\xi}   \Delta(\xi,\eta)   =0,\\
 \Delta(\xi,\eta) \equiv \pa_{\xi}\pa_{\eta} w + V^{(1)}(w).
\er
The first term in the equation above is linear in the spectral parameter $\lambda$, requiring the quantity $\Delta(\xi,\eta)$ to vanish. This condition naturally leads to the deformed SG equation of motion (\ref{sg1}). Once $\Delta(\xi,\eta)=0$ is imposed, the remaining terms in (\ref{zeroeq}) must also vanish. Consequently, the operators ${\cal L}_1  $ and ${\cal L}_2$  form a pair of linear operators associated with the deformed SG model (\ref{sg1}).
 
As a first application of the linear problem discussed above, we proceed with the construction of the energy and momentum charges.  
So, considering the identify $\pa_{\eta}(\frac{\pa_{\xi}\Phi }{\Phi}) - \pa_{\xi}(\frac{\pa_{\eta}\Phi}{\Phi})=0$ and the linear System  (\ref{lin1})-(\ref{lin2}), one has
\br
\label{ABd}
\pa_{\eta} A_{\xi} - \pa_{\xi} A_{\eta} &=& 0.
\er
So, using (\ref{AB}) one can write
\br
\label{nl0} \pa_{\eta} a_0 - \pa_{\xi} b_0 &=& 0,\\
\pa_{\eta} a_1 - \pa_{\xi} b_1 &=& 0.  \label{em1}
\er  
In fact, these eqs. define two conservation laws. The eq. (\ref{nl0}) can be written conveniently as 
\br
\pa_{\eta} [\frac{1}{3} (\pa_{\xi} w)^3] + \pa_{\xi}[V \pa_{\xi} w] = V \pa^2_{\xi} w.
\er
This is just a quasi-conservation law provided in \cite{npb1} (see the eq. (2.6) for $N=3$ of that reference).  

Whereas, the second eq. (\ref{em1}) provides the energy-momentum conservation law $\pa_{\eta} [\frac{1}{2} (\pa_
{\xi} w)^2] + \pa_{\xi} [ V-2 ] =0$, which has already been discussed in the pseudo-potential approach (\ref{n11}).

{\bf Non-local conservation laws}

A set of infinite number of non-local conservation laws for System (\ref{lin1})-(\ref{lin2}) can be constructed through an iterative procedure introduced by Br\'ezin, et.al. \cite{brezin}. In fact, this system satisfies: i) $(A_{\xi}, A_{\eta})$ is a ``pure gauge"; i.e. $A_{\mu}=\pa_{\mu}\Phi \Phi^{-1},\,\,\mu=\xi,\eta$; ii) $J_{\mu}=(A_{\xi}, A_{\eta})$ is a conserved current satisfying (\ref{ABd}). Thus, an infinite set of non-local conserved currents can be constructed using an inductive procedure. We begin by defining the currents as follows
\br
J_{\mu}^{(n)} &=& \pa_{\mu} \chi^{(n)},\,\,\,\mu\equiv \xi, \eta;\,\,\,\,n=0,1,2,...\\
d \chi^{(1)} &=& A_{\xi} d\xi + A_{\eta} d\eta ,\,\,\\
  &\equiv& d{\cal I}_0(\xi,\eta)+ \lambda\,  d{\cal I}_1(\xi,\eta),\\
J_{\mu}^{(n+1)} &=& \pa_{\mu} \chi^{(n)}-A_{\mu} \chi^{(n)},\,\,\,\,\,\chi^{(0)}=1,
\er
where
\br
\label{expan11}
d{\cal I}_0(\xi,\eta)\equiv  a_0(\xi, \eta) d\xi +b_0(\xi, \eta) d\eta ,\,\,\,\,d{\cal I}_1(\xi,\eta) \equiv a_1(\xi, \eta) d\xi +b_1(\xi, \eta) d\eta.
\er
Then one can show by an inductive procedure that the  (non-local) currents $J_{\mu}^{(n)}$ are conserved
\br
\label{nlcl}
\pa_{\mu} J^{(n)\,\mu} =0,\,\,\,\,n=1,2,3,...
\er
The first non-trivial conservation law $\pa_{\mu} J^{(1)\,\mu}=0$  reduces to the eq. (\ref{ABd}), and then provides the first two conservation laws (\ref{nl0})-(\ref{em1}). The next conservation law $\pa_{\mu} J^{(2)\,\mu}=0$ becomes
\br
\label{expan1}
\pa_{\eta}\Big[A_{\xi} - a_0 {\cal I}_0 -(a_0 {\cal I}_1 +a_1 {\cal I}_0)\lambda - a_1 {\cal I}_1 \lambda^2\Big]-\pa_{\xi}\Big[A_{\eta} -b_0 {\cal I}_0 -(b_0 {\cal I}_1 +b_1 {\cal I}_0)\lambda - b_1 {\cal I}_1 \lambda^2\Big]=0,
\er
where ${\cal I}_0$ and ${\cal I}_1$ are defined in (\ref{expan11}). From (\ref{expan1}), in addition to the conservation laws (\ref{ABd}) or  (\ref{nl0})-(\ref{em1}), one can get the next non-local conservations laws order by order in $\lambda$
\br
\pa_{\eta} (a_0 {\cal I}_0) -\pa_{\xi} (b_0 {\cal I}_0)=0,\\
\pa_{\eta}(a_0 {\cal I}_1 +a_1 {\cal I}_0) -\pa_{\xi}(b_0 {\cal I}_1 +b_1 {\cal I}_0)=0,\\ 
\pa_{\eta}( a_1 {\cal I}_1)-\pa_{\xi}( b_1 {\cal I}_1)=0.
\er 
The formulation of similar linear systems could be significant for studying deformations of well-known integrable models associated with the equation of motion (\ref{sg1}), such as the Bullough-Dodd model \cite{jhep6}.

\section{Modified NLS model as reduction of the deformed $sl(2)$ AKNS model}
 \label{sec:dnls}

The compatibility condition (\ref{compati1})-(\ref{AB11}), for special coefficients $A_j$ and $B_j$ and relevant auxiliary fields $r$ and $s$, reproduces the equation of motion of a modified AKNS system (MAKNS). A modified AKNS model has recently been studied as quasi-integrable in the anomalous zero-curvature approach \cite{paper2}. We will present a dual Riccati-type formulation and construct explicitly the first three exact conservation laws and the first two quasi-conservation laws. Moreover, a dual formulation allows us to uncover inﬁnite towers of novel anomalous conservation laws. It is shown that certain modiﬁcations of the nonlinear Schr\"{o}dinger model (MNLS) can be obtained through a reduction process starting from the MAKNS model. So, the novel inﬁnite sets of quasi-conservation laws and related anomalous charges for the standard NLS and modiﬁed MNLS cases can be constructed by an uniﬁed and rigorous approach based  on the Riccati-type pseudo-potential method. 

It is known that the standard NLS model emerges as a specific reduction of the AKNS system. Therefore, we explore a suitable deformation of the conventional pseudo-potential approach within the $sl(2)$ AKNS integrable field theory. We then examine the reduction process that leads to the modified NLS model. 

In the case of the deformed $sl(2)$ AKNS system the Riccati-type Equations (\ref{riccati11})-(\ref{riccati22}) are defined with the following quantities \cite{paper2}
\br
\label{A012nls}
A_0 &=& q,\,\,\,\, A_1 = -2 i \zeta ,\,\,\,\,A_2 = \bar{q},\\
\label{B012nls}
B_0 &=& 2 \zeta q + i \pa_x q + r,\,\,\,\,B_1 =  - 4 i \zeta^2 - i\, V^{(1)} -s,\,\,\,\,B_2 = 2 \zeta \bar{q} - i \pa_x \bar{q},\\ 
V^{(1)} & \equiv &  \frac{d V[\rho]}{d \rho},\,\,\,\rho \equiv \bar{q} q.
\er
Note that we will redefine $\xi = x,\, \eta =t$, such that $x$ and $t$ stand for space and time variables, respectively. $\pa_x q$ and $\pa_x \bar{q} $ stand for partial derivatives w.r.t. $x$. Then, one has the system of Riccati-type equations\cite{paper2}
\br
\label{ricc1nls}
\pa_x u &=& -2 i \zeta \, u + q + \, \bar{q} \,\, u^2,\\
\label{ricc2nls}
\pa_t u &=&  2 (- 2 i \zeta^2 -  \frac{1}{2} i\, V^{(1)})\, u - (-2 \zeta \bar{q} + i \pa_x \bar{q})\, u^2 + (2 \zeta q + i \pa_x q) + r - u\, s,\\
V^{(1)} &\equiv &  \frac{d V[\rho]}{d \rho},\,\,\,\rho \equiv \bar{q} q,
\er
where $u(x,t,\lambda)$ is the Riccati-type pseudo-potential, $\zeta$ is the spectral parameter and $r$ and $s$ are the auxiliary fields, and $q$ and $\bar{q}$ are the fields of the model. The potential $V(\bar{q} q)$ defines the modified AKNS equation (MAKNS). In fact, the form of (\ref{riccati11}) is similar to the first Riccati equation for the standard AKNS model. The function $B_2$ is the same as in the usual second Riccati equation; whereas, the functions $B_0$ and $B_1$ will depend on a generalized potential as compared to the standard  AKNS model \cite{prl1, wadati1}.

The auxiliary fields $r(x,t)$ and $s(x,t)$ satisfy
\br
\label{ricc1rnls}
\pa_x r &=& q \, s +  (-2 i \zeta + Q \, u)\, r,\\
\label{ricc2snls}
\pa_x s &=& Q \,r - 2  \, \bar{q}\, r + u \bar{q}\, s + 2 i X,\\
X & \equiv & - \pa_x  \( \frac{1}{2} V^{(1)} + \bar{q} q \),  \label{Xanomnls}
\er
where $Q$ is an arbitrary field. So, one has a set of two deformed Riccati-type equations for the pseudo-potential $u$ (\ref{riccati11})-(\ref{riccati22}) and a linear System (\ref{ricc1rnls})-(\ref{ricc2snls}) for the auxiliary fields $r$ and  $s$. 

Notice that, for the integrable $sl(2)$ AKNS model one has the potential \cite{paper2} 
\br
\label{nlspot}
V_{NLS}(\bar{q} q) = - \( \bar{q} q \)^2\, \rightarrow \, V_{NLS}^{(1)}(\bar{q} q) = - 2(\bar{q} q),
\er
and, therefore,  $X=0$ in (\ref{Xanomnls}), and so the auxiliary System (\ref{ricc1rnls})-(\ref{ricc2snls}) possesses the trivial solution $r=s=0$. Inserting the trivial solution into (\ref{riccati11})-(\ref{riccati22}) with the potential (\ref{nlspot}), one has a Riccati system of the standard AKNS model \cite{prl1, wadati1}.   
 
The compatibility condition   $[\pa_{t}\pa_x  u - \pa_x \pa_{t} u ]=0$ for the Riccati-type System (\ref{riccati11})-(\ref{riccati22}) with (\ref{A012nls})-(\ref{B012nls}), and taking into account  the auxiliary Equations (\ref{ricc1rnls})-(\ref{ricc2snls}), provides the eqs. of motion for the fields $q $ and $\bar{q}$
\br
\label{mnls11}
i \pa_t q + \pa^2_x q - V^{(1)} q &=& 0,\\
-i \pa_t \bar{q} + \pa^2_x \bar{q} - V^{(1)} \bar{q} &=& 0.\label{mnls22}
\er
This is a modified $sl(2)$ AKNS system (MAKNS) for arbitrary potential of type $V(\bar{q} q)$. An important observation in the constructions above is that $\frac{\pa}{\pa t} \zeta =0$, as it can be checked by direct computation using System (\ref{riccati11})-(\ref{riccati22}) and (\ref{ricc1rnls})-(\ref{ricc2snls}), provided that System (\ref{mnls11})-(\ref{mnls22}) is satisfied. So, the modified system MAKNS possesses an isospectral parameter $ \zeta $. 
 
The next identification defines the modified NLS model. So, defining \cite{paper1,proc3}  
\br
\label{nlspsi}
q \equiv  i (-\eta)^{1/2}\, \psi,\,\,\, \bar{q} \equiv  -  i (-\eta)^{1/2} \, \bar{\psi},\,\,\,\,\eta \in \IR,
\er 
in System (\ref{mnls11})-(\ref{mnls22}), with $\bar{\psi}$ being the complex conjugate of the field $\psi$, one can get the modified  NLS model
\br
\label{nlsd}
i \frac{\pa}{\pa t} \psi(x,t) + \frac{\pa^2}{\pa x^2} \psi(x,t) -  V^{(1)}(|\psi(x,t)|^2)\psi(x,t) &=&  0,\\
V^{(n)}(I)&\equiv &\frac{d^n}{d I^n} V(I),\,\,\,\, I \equiv \bar{\psi} \psi.\er 
In order to get the standard NLS model the potential and its derivatives are defined as 
\br \label{potder}
V(I)= - \eta I^2,\,\,\,V^{(1)}(I) =- 2\eta I \,\,\, \mbox{and}\,\,\,\, V^{(2)}(I)= - 2\eta.
\er 
Let us emphasize that for the standard NLS model we have the  trivial solution of System  (\ref{ricc1rnls})-(\ref{ricc2snls}), i.e. $X=0 \rightarrow r=s=0$, and the existence of the Lax pair of de ordinary NLS model reflects in its equivalent Riccati-type representation, provided by System (\ref{riccati11})-(\ref{riccati22}) with the well known potential (\ref{potder}) \cite{nucci1, nucci2, prl1, wadati1}. 

It is important to highlight that, for the standard NLS model, System (\ref{ricc1rnls})-(\ref{ricc2snls}) admits a trivial solution, namely $X=0 \rightarrow r=s=0$. Moreover, the existence of the Lax pair for the ordinary NLS model is reflected in its equivalent Riccati-type formulation, given by System (\ref{riccati11})-(\ref{riccati22}) with the well-known potential (\ref{potder}) \cite{nucci1, nucci2, prl1, wadati1}.

Next, a special space-time symmetry related to soliton-type solutions of the model is introduced. So, a reflection around a fixed point $(x_{\Delta},t_{\Delta})$ becomes
\br
\label{par1}
\widetilde{{\cal P}}:  (\widetilde{x},\widetilde{t}) \rightarrow (-\widetilde{x},-\widetilde{t});\,\,\,\,\,\,\,\,\widetilde{x} = x - x_{\Delta},\,\,\widetilde{t} = t- t_{\Delta}. 
\er 
The operator $\widetilde{{\cal P}}$ defines a shifted parity ${\cal P}_{s}$ for the variable $x$ and a delayed time reversal ${\cal T}_d$ for the variable $t$.

The standard NLS model, in the sector described by N-bright solitons possessing ${\cal C}{\cal P}_{s} {\cal T}_d$ (${\cal C} ( \psi ) \equiv\bar{\psi} $) symmetry and broken space–time translation symmetry, belongs to the family of nonlocal generalization of the NLS model considered in the recent literature \cite{ablowitz1,ablowitz2}.   

We define the quasi-integrable MAKNS model for field configurations $q$ and $\bar{q}$ that satisfy (\ref{mnls11})-(\ref{mnls22}), ensuring that the fields and the deformed potential transform under the space-time transformation (\ref{par1}) as follows
\br 
\label{aknsparity}
\widetilde{{\cal P}}(q)  = \bar{q},\,\,\,\,\,\, \widetilde{{\cal P}}(\bar{q})  = q,\,\,\,\,\,\mbox{and } \,\,\,\,\,\widetilde{{\cal P}}[ V(\rho) ] = V(\rho),\,\,\,\rho \equiv \bar{q}  q.
\er
Under this transformation, $X$ from (\ref{Xanomnls}) becomes an odd function.
\br
\label{Xtr}
\widetilde{{\cal P}}( X ) =  - X.
\er

Next, let us discuss the conservation laws in the context of the Riccati-type System  (\ref{riccati11})-(\ref{riccati22}) and the auxiliary Equations (\ref{ricc1rnls})-(\ref{ricc2snls}). So, one can get the following relationship
\br
\label{qcons2}
\pa_t [ i \bar{q} \,u  ] - \pa_x \Big[ 2 i \zeta \bar{q} \, u - \bar{q} q +  u \, \pa_x \bar{q} \Big] &=& i \bar{q} (r - s \, u). 
\er
Defining the r.h.s. of (\ref{qcons2}) as
\br
\label{qcons11}
\chi \equiv i \bar{q} (r - s \, u),
\er 
one can write a first order differential equation for the auxiliary field $\chi$ 
\br
\label{chi0}
\pa_x \chi &=& \( -2 i \zeta + 2 u \bar{q} + \frac{\pa_ x \bar{q}}{\bar{q}}\) \chi + 2 \bar{q}\, u X.
\er
Equations (\ref{qcons2}) and (\ref{chi0}) will be utilized below to reveal an infinite hierarchy of quasi-conservation laws associated with the modified AKNS model (\ref{mnls11})-(\ref{mnls22}). We will systematically construct the relevant charges, order by order, in terms of the parameter $\zeta$. Thus, we begin by considering the following expansions
\br
\label{expannls}
 u = \sum_{n=1}^{\infty} u_n \, \zeta^{-n},\,\,\,\,  \chi= \sum_{n=1}^{\infty} \chi_n  \zeta^{-n-1}.
\er
The coefficients $u_n$ can be determined order by order in powers of $\zeta$ from  the Riccati Equation (\ref{riccati11}). Likewise, using the results for the $u_n\, 's$ we get the relevant expressions for the  $\chi_n\, 's$ from (\ref{chi0}). The first components of $u_n\, 's$ and $\chi_n\,'s$ are provided in appendices A and B, respectively, of \cite{paper2}.

So, by considering the components $u_n$ and $\chi_n$ in the eq. (\ref{qcons2}) one gets the coefficient of the $n'$th order as
\br
\label{anocons}
\pa_{t} a_{x}^{(n)} &+& \pa_x a_{t}^{(n)} = \chi_{n-1},\,\,\,\,\,n=0,1,2,3,....;\, \chi_{-1} =\chi_{0}\equiv 0,\\
a_{x}^{(n)} &\equiv & i \bar{q}\, u_n ,\,\,\,\,a_{t}^{(n)} \equiv   -\( 2 i \bar{q} \, u_{n+1} - \bar{q} q \, \delta_{0, n}  + \pa_x \bar{q} \, u_{n}  \),\,\,\,\,\,u_0 \equiv 0.
\er 
Setting $\chi_{n-1} \equiv 0$ on the right-hand side of Eq. (\ref{anocons}) yields the tower of exact conservation laws characteristic of the standard AKNS system. However, when $\chi_{n}$ is nonzero, the genuinely conserved nature of this equation at each order $n$ remains to be fully understood. We will examine this construction order by order for each $\chi_{n-1}$ component. Similar quasi-conservation laws have been derived in the context of the anomalous zero-curvature formulation of the modified NLS model and its corresponding anomalous Lax pair in \cite{jhep3}.

The r.h.s. of (\ref{anocons}) for  $\chi_1, \chi_2$ and $\chi_3$ can be written as $ \chi_j \equiv \pa_{x} \chi^x_j + \pa_{t} \chi^t_j$, with $\chi^x_j$ and $\chi^t_j$ being certain local functions of $\{\bar{q}, q, V\}$ and their $x$ and  $t-$derivatives; i.e. there exist local expressions for some $\chi_j\, (j=1,2,3)$, such that the eq. (\ref{anocons}) provides a genuine local conservation law. 
 
The  {\bf zero'th} order provides a trivial identity, since $u_1=-\frac{1}{2} i q$ and $\chi_{-1}= u_0 =0$.
 
{\bf The order $n=1$ and the field normalization}

At this order the anomaly is also trivial $\chi_0 =0$ and $u_2 = \frac{1}{4} \pa_x q$. So, one has
\br
\label{n1nls}
\pa_t \( \frac{1}{2} \bar{q} q \) - \pa_x \( \frac{1}{2} i \bar{q} \pa_x q -  \frac{1}{2} i q \pa_x \bar{q}   \) =0.
\er
It provides the charge
\br
\label{n1nor}
N = \int dx \, \bar{q} q.
\er

{\bf The order $n=2$ and momentum conservation}

At this order one has $u_3 = \frac{1}{8} i (\bar{q}q q + \pa_x^2 q)$, and so
\br
\label{n2nls}
\pa_t \( \frac{1}{4} i \bar{q} \pa_x q\) + \pa_x \( \frac{1}{4} [ (\bar{q} q)^2 + \bar{q} \pa_x^2 q - \pa_x \bar{q}\pa_x q]\) &=&\chi_1.\er
The function  $\chi_1$ can be rewritten as
\br
\label{chi1}
\chi_1 &=& \frac{1}{2} \pa_x [ F(\rho) ],\,\,\,\,\rho \equiv \bar{q} q . \\
\label{fv}
F(I) & \equiv &\frac{1}{2} \rho \frac{d}{d \rho}V(\rho)-\frac{1}{2} V(\rho)+\frac{1}{2} \rho^2. 
\er
So, from (\ref{n2nls}), taking into account  (\ref{chi1}), one can write the conserved charge
\br
\label{n2mo}
P =   i \int dx \, \( \bar{q} \pa_x q - q \pa_x \bar{q} \).
\er

{\bf The order $n=3$ and energy conservation}

One has $u_4 = -\frac{1}{16}[4 \bar{q} q \pa_x q + qq \pa-x \bar{q} + \pa-x^3 q]$; so, it follows
\br
\label{n3}
\pa_t [ -\frac{1}{8} (\bar{q} q)^2 - \frac{1}{8} \bar{q} \pa^2_x q ] -\pa_x \(2 i \bar{q} u_4 + \pa_x \bar{q} u_3 \) &=& \chi_2.       
\er
The function $\chi_2$ can be rewritten as
\br
\label{chixt}
 \chi_2 \equiv  -\frac{1}{8}  \pa_t V -\frac{1}{8} \pa_t (\bar{q} q)^2 - \frac{1}{4} i \pa_x\Big[ X \bar{q} q - X (q  \pa_x \bar{q} - \bar{q}  \pa_x q)\Big].
\er
So, (\ref{n3}) provides the conserved charge
\br
\label{energy}
H_{MNLS} =   \int dx \, [\, \pa_x \bar{q} \pa_x q + V( \bar{q} q)  \,].
\er
This expression (\ref{energy})  is valid for the general MAKNS model. In particular, for the ordinary AKNS the energy follows directly  from the l.h.s. of (\ref{n3}) (provided that $\chi_2 =0 $), i.e.  $H_{NLS} =   \int dx \, [\, \pa_x \bar{q} \pa_x q + V_{NLS}( \bar{q} q)  \,]$, where $V_{NLS} = - (\bar{q} q )^2$ as in (\ref{nlspot}).

{\bf The order $n=4$: A first trivial charge and its associated anomalous charge}

One has 
\br
\label{charge4}
\pa_t \(- \frac{3}{16} i \bar{q} q \bar{q} \pa_x q -\frac{1}{32} i \pa_x (\bar{q} q)^2  - \frac{1}{16} i \bar{q} \pa^3_x q \)- \pa_x [ 2 i \bar{q} \, u_5 + \pa_x \bar{q} \, u_4 ] &=& \chi_3.
\er
Remarkably, the expression for $\chi_3$ can be written as  
\br
\label{chi3xt}
\chi_3 & \equiv & \pa_x [ \chi^{(3)}_{x} ] + \pa_t [ \chi^{(3)}_{t} ],\\
   \chi^{(3)}_{t} & = & - \frac{3}{16} i \bar{q} q \bar{q} \pa_x q  - \frac{1}{16} i \bar{q} \pa^3_x q, \label{chi3t}\\
   \label{chi3x}
    \chi^{(3)}_{x} & = & \frac{3}{8} X \bar{q} \pa_x q + \frac{3}{8}  H_1(\rho) + \frac{1}{8} \bar{q} q  \pa_x X - \frac{1}{8} X \pa_{x} (\bar{q} q) - \frac{1}{16} [\pa_{x} (\bar{q} q) ]^2+ \frac{3}{8} \bar{q} q \pa_{x}\bar{q} \pa_x q- \frac{3}{8} H_2(\rho) + \nonumber\\ 
&& \frac{3}{32} i \bar{q} \pa_t q^2 - \frac{1}{16} V^{(1)} \( q \pa^2_x \bar{q} + \bar{q} \pa^2_x q  \) + \frac{1}{16}   V^{(1)}  \pa_x q \pa_x \bar{q} +\nonumber\\ 
&& \frac{1}{16} i [\pa_t q \pa^2_x \bar{q} - \pa_x \pa_t q \pa_x \bar{q} + \bar{q} \pa_t \pa^2_x q], 
    \\   
    \frac{d}{d\rho}H_1(\rho) & \equiv & -(V^{(2)}/2+1) \rho^2,\,\,\,\,\, \frac{d}{d\rho}H_2(\rho) \equiv \rho V^{(1)},\,\,\,\,\,\rho \equiv \bar{q}  q.
\er
After a careful examination of the eq. (\ref{charge4}) one can argue that the charge $Q^{(4)}$ trivially vanishes.

However, at this and higher orders, an asymptotically conserved charge can be defined for the MAKNS model. In this case one has
\br
\label{charge4a}
Q^{(4)}_a =  \frac{i}{2} \int dx \, \Big[ 3 \bar{q}  q \( \bar{q} \pa_x q - q \pa_x \bar{q} \) + \bar{q} \pa^3_x q - q \pa^3_x \bar{q}  \Big],
\er
such that 
\br
\label{qa44}
\frac{d}{dt}  Q^{(4)}_a     &=& - 8 \int\, dx \, \chi_3,\\
\label{qa442}
                 &=&  -  \int\, dx \, [ 3(\bar{q} q)^2 X + \bar{q} q \pa^2_x X - 3 \pa_x \bar{q} \pa_x q X].
\er
Notice that the anomaly density in (\ref{qa442}) possesses an odd parity under (\ref{par1}) and (\ref{aknsparity}) taking into account that $X$ is an odd function according to (\ref{Xtr}). Therefore, one has $\int dt \int dx\, \chi_3 = 0$ implying the asymptotically conservation of the charge $Q^{(4)}_a$, i.e. $Q^{(4)}_a(+ \infty)=Q^{(4)}_a(-\infty)$.   

The charge $Q^{(4)}_{a}$ in (\ref{charge4a}) has the same form as the fourth-order charge in the standard AKNS model. Specifically, when the right-hand side of (\ref{charge4}) vanishes, i.e., $\chi_3 =0$, the resulting charge takes a form similar to (\ref{charge4a}), suitably rewritten by discarding surface terms. Considering the reduction process in (\ref{nlspsi}), a corresponding anomalous charge for the MNLS model can be obtained, as discussed in \cite{jhep3, jhep4, proc1, jhep5}. In fact, under the reduction (\ref{nlspsi}), the anomalous charge  $Q^{(4)}_{a}$ in (\ref{charge4a}) corresponds to the one derived for the MNLS model in Section 3.5 of \cite{paper1,proc3}.

{\bf The order $n=5$ and the quasi-conserved charge}

At this order one can write \cite{paper2}
\br
\label{charge5nls}
 \frac{1}{32} \pa_t  \Big[ 2 (\bar{q} q)^3  + 5 \bar{q}^2 (\pa_x q)^2 + 6 \bar{q}q \( \pa_x q \pa_x \bar{q} + \bar{q}  \pa^2_x q\) + \bar{q} q^2 \pa^2_x \bar{q} + \bar{q} \pa^4_x q  \Big]- \pa_x [ 2 i \bar{q} \, u_6 + \pa_x \bar{q} \, u_5 ] = \chi_4,
\er
with
\br
\label{chi4xt}
\chi_4 & \equiv & \pa_x [ \chi^{(4)}_{x} ] +\frac{1}{16} \pa_t [ Z(\rho)] +\beta_1,\\
\label{anomaly1}
\beta_1 &=&  \frac{i}{32} \( \frac{2 \bar{q} q}{V^{(1)} } + 1\)  \( \bar{q} \pa_{x}^4 q - q \pa^4_{x} \bar{q}  \) V^{(1)}
     \\     
 \label{zz1}   \frac{d}{d\rho} Z(\rho) & \equiv &   6 \int_{\rho_0}^{\rho} \hat{\rho} [ \frac{1}{2} V^{(2)}(\hat{\rho}) + 1 ] \, d\hat{\rho},
\er
where $\chi^{(4)}$ is provided in eq. (2.44)-(2.46) of \cite{paper2}.

One can define the fifth order quasi-conservation law as 
\br
 \label{charge5a}
\frac{d}{dt} Q^{(5)}_a &=&  \int dx\, \beta_1,\\
Q^{(5)}_a & \equiv & \frac{1}{32} \int dx\,  \Big[ 2 (\bar{q} q)^3  - 8 \bar{q}q \pa_x q \pa_x \bar{q} - \bar{q}^2   (\pa_x q)^2 - q^2 (\pa_x \bar{q})^2 + \pa_x^2 \bar{q} \pa_x^2 q - 2 Z(\rho) \Big] \label{charge5a1}.
\er
Notice that the anomaly in (\ref{anomaly1}) possesses an odd parity under (\ref{par1}) and  (\ref{aknsparity}). Then, one has $\int dt \int dx \, \beta_1 = 0$ implying the asymptotically conservation of the charge $Q^{(5)}_a$ ($Q^{(5)}_a(+\infty)=Q^{(5)}_a(-\infty)$).   

Notice that, in the standard AKNS  limit, i.e. when $V^{(1)} = - 2 \bar{q} q$ and $V^{(2)} = - 2$ for the AKNS potential (\ref{nlspot}), the factor $\(\frac{2 \bar{q} q}{V^{(1)}} + 1\)$ of the anomaly $\beta_1$ in (\ref{anomaly1}) vanishes, and the term $Z(\rho)$ in (\ref{charge5a1}) can be set to zero (see (\ref{zz1})). So, the quasi-conserved charge $Q^{(5)}_a$ in (\ref{charge5a}) becomes the fifth order charge $Q^{(5)}$  of the usual AKNS model. Actually, in this limit one has that $\chi_4 =0$ for $X=0$, then the r.h.s. of (\ref{charge5nls}) vanishes, and so, this eq. is as an exact conservation law.

It has been constructed the set of (quasi-)conservation laws of type (\ref{anocons}) using the Riccati-type approach of the modified AKNS model. Through a suitable reduction process, these charges correspond to those of the MNLS model. In \cite{jhep3}, within the anomalous Lax pair framework of modified NLS models and via the abelianization procedure, an infinite set of asymptotically conserved charges was derived, resembling the exact conserved charges of the standard NLS model.

\subsection{Dual Riccati-type formulation and novel anomalous charges} 

In order to discuss a dual formulation, let us consider the system formed by the Riccati-type eqs. (\ref{ricc1nls})-(\ref{ricc2nls}) and the auxiliary eq. (\ref{chi0})  
\br
\label{ricc1n1}
\pa_x u &=& -2 i \zeta \, u + q + \, \bar{q} \,\, u^2,\\
\label{ricc2n1}
\pa_t u &=&  2 (- 2 i \zeta^2 -  \frac{1}{2} i\, V^{(1)})\, u - (-2 \zeta \bar{q} + i \pa_x \bar{q})\, u^2 + (2 \zeta q + i \pa_x q) - i \frac{\chi}{\bar{q}},\\
\pa_x \chi &=& \( -2 i \zeta + 2 u \bar{q} + \frac{\pa_ x \bar{q}}{\bar{q}}\) \chi + 2 \bar{q}\, u X \label{chi01}.
\er
Note that the system of differential eqs. of the AKNS model (\ref{mnls11})-(\ref{mnls22}) is invariant under the transformations: $q  \leftrightarrow \bar{q}$ and $i \leftrightarrow -i$.

A dual formulation of the Riccati-type System (\ref{ricc1n1})-(\ref{chi01}) is achieved by performing the changes $q  \leftrightarrow \bar{q}$ and $i \leftrightarrow -i$, $u \rightarrow \bar{u}$ and $\chi \rightarrow \bar{\chi}$ into System (\ref{ricc1n1})-(\ref{ricc2n1}) and (\ref{chi01}). 

So, one has a dual Riccati-type system \cite{paper2}
\br
\label{ricc1d}
\pa_x \bar{u} &=& 2 i \zeta \, \bar{u} + \bar{q} + \, q \,\, \bar{u}^2,\\
\label{ricc2d}
\pa_t \bar{u} &=&  2 (2 i \zeta^2 +  \frac{1}{2} i\, V^{(1)})\, \bar{u} + (2 \zeta q + i \pa_x q)\, \bar{u}^2 + (2 \zeta \bar{q} - i \pa_x \bar{q})+i \frac{\bar{\chi}}{q},\\
\label{chi0d}
\pa_x \bar{\chi} &=& \( 2 i \zeta + 2 \bar{u} q + \frac{\pa_ x q}{q}\) \bar{\chi} + 2 q\, \bar{u} X,
\er
where $\bar{u}$ is a new Riccati-type pseudo-potential and $\bar{\chi}$ is a new  auxiliary field. Notice that $X$ defined in (\ref{Xanomnls}) is an invariant. It is a simple calculation to verify that this dual Riccati-type System (\ref{ricc1d})-(\ref{chi0d}) reproduces the AKNS Equations (\ref{mnls11})-(\ref{mnls22}).  

Consider the expansions
\br
\label{expandual}
\bar{u} = \sum_{n=1}^{\infty} \bar{u}_n \, \zeta^{-n},\,\,\,\,  \bar{\chi}= \sum_{n=1}^{\infty} \bar{\chi}_n  \zeta^{-n-1},
\er
where the coefficients $\bar{u}_n$ and $\bar{\chi}_n\, 's$ can be obtained from (\ref{ricc1d}) and (\ref{chi0d}), respectively.  The first components  are provided in appendix C of \cite{paper2}.

For the fields $q, \bar{q}$ and $X$ whic satisfy (\ref{aknsparity}) and (\ref{Xtr}), respectively, one can verify from (\ref{ricc1n1})-(\ref{chi01}) and (\ref{ricc1d})-(\ref{chi0d})  the next symmetry transformations 
\br
\label{ubu}
\widetilde{{\cal P}}(u) &=& - \bar{u},\,\,\,\,\,\,\,\widetilde{{\cal P}}(\bar{u}) = - u,\\
\widetilde{{\cal P}}(\chi) &=&- \bar{\chi},\,\,\,\,\,\,\, \widetilde{{\cal P}}(\bar{\chi})= - \chi.
\label{chibchi}
\er 
In fact, by inspecting the first few components in appendix C of \cite{paper2}  one can show
\br
\label{ubun}
\widetilde{{\cal P}}(u_n) &=& - \bar{u}_n,\,\,\,\,\,\,\,\widetilde{{\cal P}}(\bar{u}_n) = - u_n,\,\,\,\,\,\, n=1,2,...,6,\\
\widetilde{{\cal P}}(\chi_n \pm  \bar{\chi}_n) &=& \mp  (\chi_n \pm \bar{\chi}_n),\,\,\,\,\,\,\,\,\,\,\,\,\,\,\,\,\,\,\,\,\,\,\,\,\,\,\,\,\,\,\,\,\,n=1,2,...,5.
\label{chibchin}
\er  
For the both dual systems of Riccati-type eqs. (\ref{ricc1n1})-(\ref{chi01}) and (\ref{ricc1d})-(\ref{chi0d}) one can write the next equations, respectively
\br
\label{qua1d}
\pa_t(i \bar{q} u) -\pa_x (2i \zeta \bar{q} u - \bar{q}q+ u \pa_x \bar{q}) = \chi,
\er
and 
\br
\label{qua2d}
\pa_t(i q \bar{u}) -\pa_x (2i \zeta q \bar{u} + \bar{q}q - \bar{u} \pa_x q) = -\bar{\chi},
 \er
where (\ref{qua1d}) has already been expressed in (\ref{qcons2}) with $\chi$ defined in (\ref{qcons11}). Subtracting  the b.h.s. of (\ref{qua1d}) and (\ref{qua2d}) one has 
\br
\label{quasidual}
\pa_t [i \bar{q} u-i q \bar{u}] -\pa_x [2i \zeta (\bar{q} u -q \bar{u})- 2\bar{q}q+ u \pa_x \bar{q} + \bar{u} \pa_x q] = \chi + \bar{\chi}.
\er 

Observe that the right-hand side of the last equation is an odd function under the special space-time operator. Consequently, Equation (\ref{quasidual}) represents a quasi-conservation law. Indeed, the first five lowest-order components are explicitly odd functions, as shown in (\ref{chibchin}).

A standard computation reveals that the expansion of the quasi-conservation law (\ref{quasidual}) in powers of $\zeta^{-n}$ yields the conserved charges associated with normalization $(n=1)$, momentum$(n=2)$, and energy $(n=3)$. Higher-order terms, on the other hand, produce the same anomalous charges discussed in Section \ref{sec:riccati}.

Notably, the charge densities in the (quasi-)conservation Equation (\ref{quasidual}) are even functions. This follows from the fact that the expression $[i \bar{q} u-i q \bar{u}]$  within the partial time derivative on the left-hand side of (\ref{quasidual}) has even parity.

However, the summation of the b.h.s. of (\ref{qua1d}) and (\ref{qua2d}) will not reproduce a quasi-conservation law,  since the anomaly $(\chi - \bar{\chi})$ is an even function according to  (\ref{chibchi}). In addition, in this case the expression $[i \bar{q} u+i q \bar{u}]$ of the charge density  will be an odd parity function, furnishing a trivial charge.

It is possible to construct new towers of quasi-conservation laws with true anomalies, meaning anomaly densities that exhibit odd parity under the symmetry transformation (\ref{par1}). To achieve this, one must seek conservation equations with charge densities of even parity, such as $ \bar{u}u$, $i(\bar{u}\pa_x u- u \pa_x\bar{u})$,  $ \pa_x \bar{u} \pa_x u$, $ i(\bar{u}\pa_x^3 u- u \pa_x^3\bar{u})$, $i\bar{u} u (\bar{u}\pa_x u- u \pa_x\bar{u}),...$, among others.

Following this approach, an infinite hierarchy of new quasi-conservation laws has been constructed in \cite{paper2}, each building upon monomials or polynomials of these forms. In \cite{paper1,proc3}, an infinite set of anomalous charges for the modified NLS equation was derived through a direct construction method, while \cite{paper2} uncovered novel anomalous charges via the pseudo-potential approach.

These findings undoubtedly merit a more thorough investigation within the framework of (quasi-)integrability phenomena and, more broadly, in the context of soliton collision dynamics—an avenue that warrants further exploration in future work.

Moreover, formulating the modified AKNS System (\ref{mnls11})–(\ref{mnls22}) as a linear system enables the construction of an infinite set of non-local conserved charges \cite{paper2}.

AKNS-type models are widespread in nonlinear science, making it worthwhile to explore the significance and physical implications of the infinite towers of anomalous and non-local charges. In recent years, remarkable and profound connections between integrable models and gauge theories have been uncovered. Notably, a form of triality has been proposed, linking gauge theories, integrable models, and gravity theories in certain UV regimes. In particular, the $(1+1)D$ nonlinear Schrödinger equation corresponds to the $2D\, {\cal N} = (2, 2)^{\star} U(N )$  super Yang-Mills theory (see \cite{nian} and references therein).

\section{Riccati-type pseudo-potentials and deformations of KdV}
\label{sec:dkdv}

A specific deformation of the KdV model has been analyzed in \cite{npb} within the framework of the anomalous zero-curvature formulation. In connection with this, an alternative scheme involving a direct deformation of the model’s Hamiltonian structure was proposed in \cite{abhinav2}. In the present work, we demonstrate that both of these frameworks can be systematically embedded within the unified Riccati-type pseudo-potential formalism developed in \cite{jhep33, proc2}, wherein an infinite hierarchy of asymptotically conserved charges emerges.

A deformed KdV system can be defined as the Riccati-type Equations (\ref{riccati11})-(\ref{riccati22}) with the following quantities \cite{jhep33, proc2}
\br
\label{A012kdv}
A_0 &=& U,\,\,\,\, A_1 = -2 \lambda,\,\,\,\,A_2 = 2,\,\,\,
B_0 = -4 \lambda^2 U - 4 U^2 +2 \lambda U_x - U_{xx}+Y+\chi, \\ 
\label{B012kdv}
B_1 &=& 2 (4 \lambda^3 + 4 \lambda U - 2 
U_x ),\,\,\,\,B_2 = - 8( \lambda^2 +  U),
\er
where the notation $U_x,\, U_{xx}$ stand for $\frac{\pa U}{\pa x}, \frac{\pa^2 U}{\pa x^2}$ and the space and time coordinates identifications $\xi =x,\, \eta =t$ have been made. Then, one has the system of Riccati-type equations\cite{jhep33} 
\br
\label{r1}u_x &=& U + 2 u^2 - 2 \lambda u,\\
\label{r2}
u_t&=& -4 \lambda^2 U - 4 U^2- 8( \lambda^2 +  U) u^2  +2 u (4 \lambda^3 + 4 \lambda U - 2 
U_x )+2 \lambda 
U_x - U_{xx}+Y+\chi,
\er
where the field $U(x,t,)$ is a KdV type field, $Y(x,t)$ encodes the deformation away from the KdV model, $u(x,t,\lambda)$ is a Riccati-type pseudo-potential and  $\chi(x,t,\lambda)$ is an auxiliary field satisfying
\br
\label{aux1}
\pa_x \chi +  2 (\lambda - 2 u) \chi = -2 (\lambda - 2 u) Y.
\er
In the equations above $\lambda$ plays the role of a spectral parameter. The eq. (\ref{aux1}) is a non-homogeneous ordinary differential equation for $\chi$ in the variable $x$, which can be  integrated  by quadratures.  

Next, the compatibility condition (\ref{compati1})-(\ref{AB11}) for System (\ref{r1})-(\ref{r2}), i.e. $(\pa_t \pa_x u - \pa_x \pa_t u)=0$ together with the auxiliary eq. (\ref{aux1}) furnishes the equation
\br
\label{dkdv2}
U_t + 12 U U_x + U_{xxx} = Y_x.
\er
This is a deformed KdV equation. Notice that for $Y=0$ and $\chi =0$ one has the Riccati System (\ref{r1})-(\ref{r2}) approach for the KdV equation
\br
U_t + 12 U U_x + U_{xxx} = 0.
\er 
Since $Y$ can be regarded as an arbitrary functional of $U$ and its derivatives-encompassing both local and non-local terms, as well as potential deformation parameters-Equation (\ref{dkdv2}) represents a generalized deformation of the KdV model within the pseudo-potential framework. This deformation is characterized by a set of parameters $\{\epsilon_i\}$, where setting $\epsilon_i =0$ recovers the standard KdV model. Thus, the problem reduces to determining the existence of the auxiliary field $\chi$.

In order to define quasi-integrability in deformed KdV models it is convenient to introduce some space-time symmetries related to soliton-type solutions. So, consider the space-time reflection around a fixed point $(x_{\Delta},t_{\Delta})$
\br
\label{parity1}
{\cal P}:  (\widetilde{x},\widetilde{t}) \rightarrow (-\widetilde{x},-\widetilde{t});\,\,\,\,\,\,\,\,\widetilde{x} = x - x_{\Delta},\,\,\widetilde{t} = t- t_{\Delta}. 
\er 
The transformation ${\cal P}$ defines a shifted parity ${\cal P}_{s}$ for the $x$ variable and the delayed time reversal ${\cal T}_d$ for the $t$ variable. For the type of deformations $Y$ satisfying 
\br
\label{parity11}
{\cal P}(U) &=& U,\\
{\cal P}(Y) &=& Y,
\label{parity22}
\er
the deformed KdV model (\ref{dkdv2}) will be considered quasi-integrable.   
   
A notable feature of the Riccati-type approach outlined above is its capacity to incorporate more general deformations. In this framework, the field $Y$ can depend on arbitrary functions of $U$ and its derivatives, as well as on auxiliary fields. Among non-local deformations, an interesting avenue for exploration is the recently introduced Alice-Bob (AB) physics \cite{alice1, alice2}, particularly in the context of quasi-integrability.

Next, we analyze the anomalous conservation laws within the pseudo-potential framework. To this end, we consider the relevant (quasi-)conservation law expressed in terms of the pseudo-potential field $u$ and the auxiliary field $\chi$. From (\ref{r1})–(\ref{r2}), one can write the following equation\footnote{Various expressions of this form can be constructed; here, we adopt a formulation where the non-homogeneous terms on the right-hand side explicitly involve the deformation variable $\chi$. This ensures that when $\chi=0 (Y=0)$, the left-hand side reconstructs, order by order in $\lambda^{-1}$, the standard conservation laws of the usual KdV model.}
\br
\label{rr1}
\frac{\pa}{\pa t} u + \frac{\pa}{\pa x}( 4 \lambda^2 u + 4 U u - 2 \lambda U + U_x) = Y + \chi.
\er 
From this point onward, we derive the relevant (quasi-)conservation laws in terms of the fields of the deformed KdV model. To this end, we consider expansions in powers of the $\lambda$ parameter
\br
\label{pow1}
u &=& \sum_{n=0} c_n \lambda^{-n-1},\\
\label{pow2}
\chi &=& \sum_{n=1} d_n \lambda^{-n+1}.
\er 
Substituting these expansions into (\ref{r1}) and (\ref{aux1}), the first $c_n$'s and $d_n$'s take the form \cite{jhep33, proc2}
\br
\nonumber
c_0 &=& \frac{1}{2} U,\\
\nonumber
c_1 &=& -\frac{1}{4} U_x,\\
\nonumber
c_2 &=& \frac{1}{4} (U^2 +\frac{1}{2} U_{xx}),\\
\label{cc11}
c_3 &=& -\frac{1}{2} (U U_x + \frac{1}{8} U_{xxx}),\\
\nonumber
c_4 &=&  \frac{1}{32} (8 U^3+10 U_x^2+12 U U_{xx}+U_{xxxx}),\\
\nonumber
c_5 &=& - \frac{1}{64} (64 U^2 U_x + 36 U_x U_{xx} + 16 U U_{xxx} + U_{xxxxx}).
\er 
and
\br
\nonumber
d_1&=& -Y,\\
\nonumber
d_2 &=& \frac{1}{2} Y_x,\\
\nonumber
d_3 &=& -\frac{1}{4} Y_{xx},\\
\label{dd11}
d_4 &=& \frac{1}{8} (4 U Y_x +Y_{xxx}),\\
\nonumber
d_5 &=& -\frac{1}{16} (8 U_x Y_x+ 8 U Y_{xx} + Y_{xxxx}),\\
\nonumber
d_6 &=& \frac{1}{32} (24 U^2 Y_x +12 U_{xx} Y_x + 20 U_x Y_{xx}+ 12 U Y_{xxx} + Y_{xxxxx}).
\er 
Notice that by setting $Y=0$ and $\chi =0$ on the right-hand side of Eq. (\ref{rr1}), it transforms into a strictly exact conservation law. This allows for the construction of an infinite hierarchy of exact conservation laws for the standard KdV equation. Next, by utilizing the power series expansions of $u$ (\ref{pow1}) and $\chi$ (\ref{pow2}) in terms of $\lambda^{-1}$, and substituting them into Eq. (\ref{rr1}), one obtains a polynomial in powers of $\lambda^{-1}$ for $c_n$'s and $d_n$'s. Considering the expressions for $c_n$'s and $d_n$'s, it follows that the first two terms in this series ($n=-1,0$) yield trivial equations. Similarly, for  $n\geq 1$, one can write
\br
\label{setr}
\pa_t \( c_{n-1} \) + \pa_x \(4c_{n+1} + 4 U c_{n-1}\) = d_{n+1},\,\,\,\,\,n=1,2,3....
\er
This is an infinite set of quasi-conservation laws for the deformed KdV model (\ref{dkdv2}) in the Riccati-type pseudo-potential approach.

A careful examination of the conservation laws corresponding to even-order terms in $\lambda^{-n},\,n= 2, 4, 6$ reveals that these equations are merely the $x-$derivatives of the conservation laws associated with $n=1,3,5$, respectively. Consequently, they do not introduce any new conservation laws. Then, we consider below the first three non-trivial cases for $n=1,3,5$.

{\bf First order} ($n=1$)
\br
\label{n1kdv}
\pa_t \(\frac{1}{2} U\) - \pa_x \( 4c_2 + 4 U c_0 - \frac{2}{2} Y\) =0.
\er
Then, taking into account (\ref{tr0}) one can define the charge
\br
\label{mass}
q^{(1)} = \frac{\a}{12} \int_{-\infty}^{+\infty} dx \, U.  
\er
This charge is typically linked to the ``mass". Because it arises from an exact conservation law, it remains conserved even in the deformed KdV model (\ref{mrkdv}). 

{\bf Third order} ($n=3$)

\br
\label{n33}
\pa_t [\frac{1}{4} U^2 + \frac{1}{3} U_{xx}] + \pa_x (4 c_3 + U c_2) = \frac{1}{2} U Y_x + \frac{1}{8} \pa_x (Y_{xx}).
\er
A remarkable fact is that the r.h.s. of (\ref{n33}), using the eq. of motion (\ref{dkdv2}), can be written as 
\br
\frac{1}{2} U Y_x + \frac{1}{8} \pa_x (Y_{xx}) =
 \pa_t (\frac{1}{4} U^2) + \pa_x [ 2 U^3 + \frac{1}{2} U U_{xx}-\frac{1}{4} (U_x)^2].\label{xm3}
\er
Substituting the last identity into Eq. (\ref{n3}) yields a trivial identity, resulting in a vanishing charge density. Consequently, the charge at this order is trivial
\br
q^{(3)} =0.\label{q3}
\er
However, following the developments in \cite{npb, jhep33, proc2},  at this order one can define the quasi-conservation law
\br
\label{qc3}
\frac{d q^{(3)}_a}{dt} = \a^{(3)},
\er
where
\br
\label{qan1}
q^{(3)}_a \equiv  \frac{1}{4} \int_{-\infty}^{+\infty} dx \, U^2 ,\,\,\,\,\,\,\, \a^{(3)}  \equiv   \int_{-\infty}^{+\infty} dx  \, \frac{1}{2} U Y_x,
\er 
is the asymptotically conserved charge $q_a^{(3)}$, with $ \a^{(3)}$ being its relevant anomaly. The space-time integral $\int_{-\infty}^{+\infty} dt\, \a^{(3)} $ vanishes for field configurations satisfying the symmetry (\ref{parity1})-(\ref{parity22}), rendering the charge $q_a^{(3)}$ to be an asymptotically conserved charge, i.e. $q_a^{(3)}(-\infty) = q_a^{(3)} (+\infty).$

{\bf Fifth order}($n=5$)

At this order one can write
\br
\label{n5}
\pa_t [\frac{1}{4} U^3-\frac{1}{16} U_{x}^2 + \pa_x(\frac{3}{8} UU_x + \frac{1}{32} U_{xxx})] + \pa_x (4 c_6 + 4U c_4) &=& \pa_t(\frac{1}{4} U^3 -\frac{1}{16} U_x^2) + \pa_x[\frac{1}{8} U_x U_t + \nonumber \\
&&\frac{3}{4} U^2 U_{xx} + \frac{1}{16} U_{xx}^2].
\er
Therefore, one has a trivially vanishing charge also at this order. Following \cite{npb, jhep33, proc2} one can define the quasi-conservation law
\br
\frac{d q^{(5)}_a}{dt} = \a^{(5)},
\er
where
\br
\label{qan2}
q^{(5)}_a \equiv  \int_{-\infty}^{+\infty} dx  [ \frac{1}{4} U^3-\frac{1}{16} U_{x}^2],\,\,\, \a^{(5)}  \equiv   \int_{-\infty}^{+\infty} dx  \, (\frac{3}{4} U^2 + \frac{1}{8} U_{xx}) Y_x,
\er 
are the asymptotically conserved charge $q^{(5)}_a$ and its relevant anomaly $ \a^{(5)}$. This charge maintains the same form as in the usual KdV at this order. Similarly, as above the space-time integral $\int_{-\infty}^{+\infty} dt\, \a^{(5)} $ vanishes for solutions with the symmetry (\ref{parity1})-(\ref{parity22}), such that the charge $q_a^{(5)}$ becomes an asymptotically conserved charge, i.e. $q_a^{(5)}(-\infty) = q_a^{(5)} (+\infty).$

Similar construction can be performed for the higher order asymptotically conserved charges. Note that in the ordinary KdV, i.e. when the anomaly $Y=0$,   the  charges  of (\ref{mass}), (\ref{qan1}) and (\ref{qan2}) are usually associated with the
``mass", ``momentum" and ``energy" conservation, respectively. 

\subsection{Particular deformations of the KdV}

A particular deformation of the KdV model has been considered \cite{npb, jhep33, proc2} by making the identification
\br
\label{tr0}
U &\equiv& \frac{\a}{12} V + \frac{\a}{144},\\
Y &=&-\frac{\a}{12} [\frac{\a}{4} \epsilon_2 w_x v_t -\frac{12 \epsilon_1}{\a} (V_{xt}+V_{xx})], \label{tr1}
\er
where $Y$ is defined such that the auxiliary fields satisfy 
\br
\label{wv1}
V &=& w_ t,\\
V &=& v_x . \label{wv2}
\er
So, substituting into (\ref{dkdv2}) one gets the equation \cite{npb, jhep33, proc2}
\br
\label{mrkdv}
V_t + V_x +\Big[\frac{\a}{2} V^2 + \epsilon_2 \frac{\a}{4} w_x v_t + V_{xx} - \epsilon_1 (V_{xt} + V_{xx} )\Big]_x
= 0.
\er
This is a deformation of the KdV model. Notice that for  $\epsilon_1= \epsilon_2=0$ one recovers the usual KdV model. The real parameters $\epsilon_1$  and  $\epsilon_2$ serve as deformation parameters that modify the standard KdV equation, while  $\a$ is an arbitrary real parameter. The formalism imposes no constraint on the magnitudes of the deformation parameters $\{\epsilon_1, \epsilon_2\}$; they may take arbitrary values without compromising validity. The soliton-like behavior in these deformed models arises from the underlying quasi-integrable structure rather than perturbative expansions around small deformation parameters. The soliton-like behavior should be contrasted with that of genuine solitons, which preserve both their velocity and spatial profile following interactions, with the only measurable consequence of the collision being a phase shift.

The model (\ref{mrkdv}) encompasses various sub-models. Specifically, when $\epsilon_1=  \epsilon_2=0$, the equation reduces to the integrable KdV model. Setting  $\epsilon_1= 1,  \epsilon_2=0$ corresponds to the regularized long wave (RLW) equation, which is non-integrable but admits a one-soliton solution. Although two- and three-soliton solutions for the RLW model have been constructed numerically, their analytical forms remain unknown.

In contrast, when $\epsilon_1= \epsilon_2=1$, the equation describes the modified regularized long wave (mRLW) equation, which exhibits the notable property of possessing exact analytical two-soliton solutions. Furthermore, for $\epsilon_2=0,\,\epsilon \neq \{0,1\}$, the model corresponds to the KdV-RLW or Korteweg–de Vries–Benjamin–Bona–Mahony (KdV-BBM) type equations. 
 
So, the Riccati-type pseudo-potential approach includes the model (\ref{mrkdv}) considered in \cite{npb, jhep33,proc2} and it is suitable to explore a deformation of the KdV equation which includes non-local terms such as $Y$ in (\ref{tr1}).

In order to illustrate the vanishing anomaly property and the asymptotically conserved charge let us take the third order quasi-conservation law (\ref{n33}) and its associated asymptotically conserved charge $q^{(3)}_a$ in (\ref{qc3}), adapted to the deformed model (\ref{mrkdv}). So, one must take into account the identifications (\ref{tr0})-(\ref{tr1}) and (\ref{wv1})-(\ref{wv2}). So, let us take the time integral of the quasi-conservation law (\ref{qc3}) as
\br
\label{qc3t1}
\int_{-\infty}^{+\infty} dt  \frac{d q^{(3)}_a}{dt} &=& \int_{-\infty}^{+\infty} dt   \a^{(3)},\\
\label{qc3t2}
&=&  \int_{-\infty}^{+\infty} dt \int_{-\infty}^{+\infty} dx  \, \frac{1}{2} U Y_x\\
&=& - \int_{-\infty}^{+\infty} dt \int_{-\infty}^{+\infty} dx  \, \frac{1}{2} U_x Y,\label{qc3t3}
\er
where the line (\ref{qc3t3}) is achieved after an integration by parts and assuming the boundary condition $U(|x|= +\infty) =0$. Next, taking into account (\ref{tr0})-(\ref{tr1}) the last identity (\ref{qc3t3}) can be written as
\br
\label{xtano10}
q^{(3)}_a(t=+\infty)-q^{(3)}_a(t=-\infty) &=&  \int_{-\infty}^{+\infty} dt \int_{-\infty}^{+\infty} dx \Big[ \frac{\a^3}{3^2 2^7} \epsilon_2 V_x w_x v_t - \frac{\a}{3 2^4} \epsilon_1 (\pa_t V_x^2 + \pa_x V_x^2 )\Big]\\
&=& \frac{\a^3 \epsilon_2}{3^2 2^7} \int_{-\infty}^{+\infty} dt \int_{-\infty}^{+\infty} dx\, V_x w_x v_t, \label{xtano1}
\er 
where the last two terms in the r.h.s. of (\ref{xtano10}), being total derivatives with respect to space and time, respectively, vanish upon the integral evaluation under the assumptions $V_x^2 (|x|=\infty,t)= 0,\, V_x^2 (x,|t|=\infty)= 0$. 
  
Next, let us analyze the symmetries of the fields. The field configurations satisfying the space-time symmetry (\ref{parity1}) with even parity in (\ref{parity22}), for the model (\ref{mrkdv}) implies
\br
\label{parityV1}
{\cal P}(V)= V,\,\,\,\,
{\cal P}(w) = - w,\,\,\,\,\, {\cal P}(v) = - v.\er
Since the integrand in equation (\ref{xtano1}) is odd under the transformation (\ref{parityV1}), the space-time integral on the right-hand side of (\ref{xtano1}) vanishes when evaluated over the entire space-time domain
\br
\label{vanish}
\int_{-\infty}^{+\infty} dt \int_{-\infty}^{+\infty} dx\, V_x w_x v_t =0.
\er
So, takinkg into account (\ref{vanish}), from (\ref{xtano1}) one can write
\br
\label{asym3a}
q^{(3)}_a(t=+\infty) = q^{(3)}_a(t=-\infty).
\er   
This demonstrates that the charge $q^{(3)}_a$ is asymptotically conserved, attaining the same constant value in the remote past and far future, despite potential variations during the intermediate interaction. Below, we will numerically verify that $q^{(3)}_a$   remains effectively constant throughout the entire two-soliton scattering process, up to numerical precision. 
\begin{figure}
\centering
\label{fig2}
\includegraphics[width=2cm,scale=6, angle=0,height=8cm]{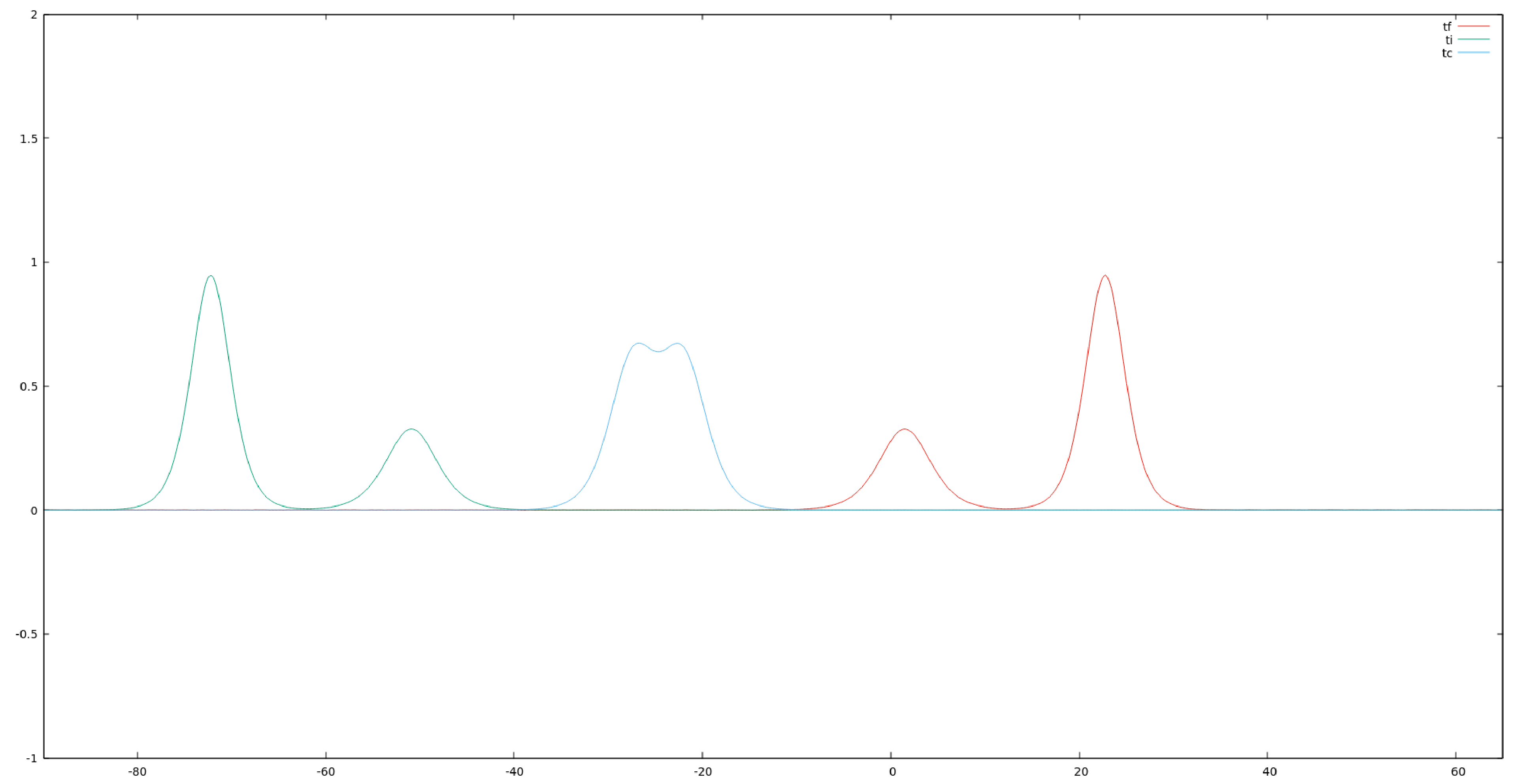} 
%\begin{center}
\parbox{6in}{\caption{(color online) Numerical simulation of 2-soliton collision of the model (\ref{mrkdv}) for three successive times, $t_i$, before collision (green); $t_c$, collision (blue) and $t_f$, after collision (red). The parameter values are $\epsilon_1 = 1.2, \epsilon_2 = 0.9, \alpha = 4$.}}
%\end{center}
\end{figure}

\begin{figure}
\centering
\includegraphics[scale=0.25]{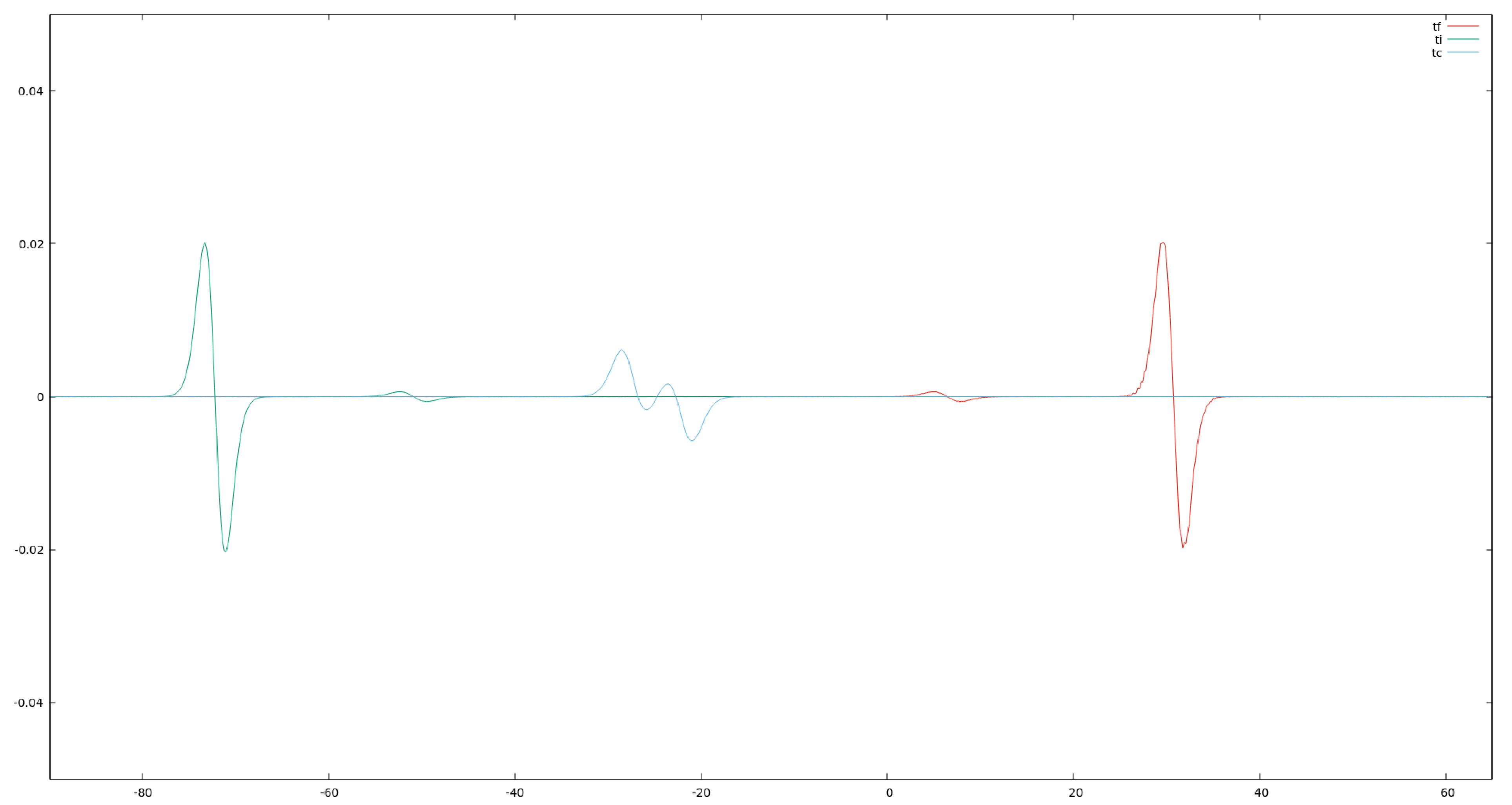}\\
\includegraphics[scale=0.25]{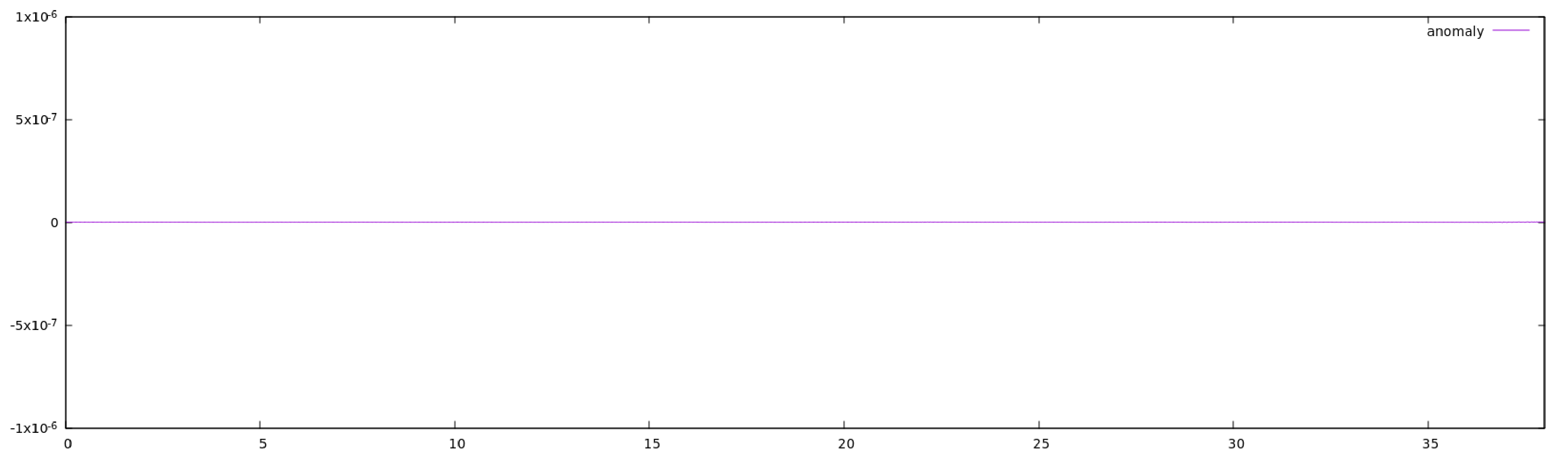}
\includegraphics[scale=0.25]{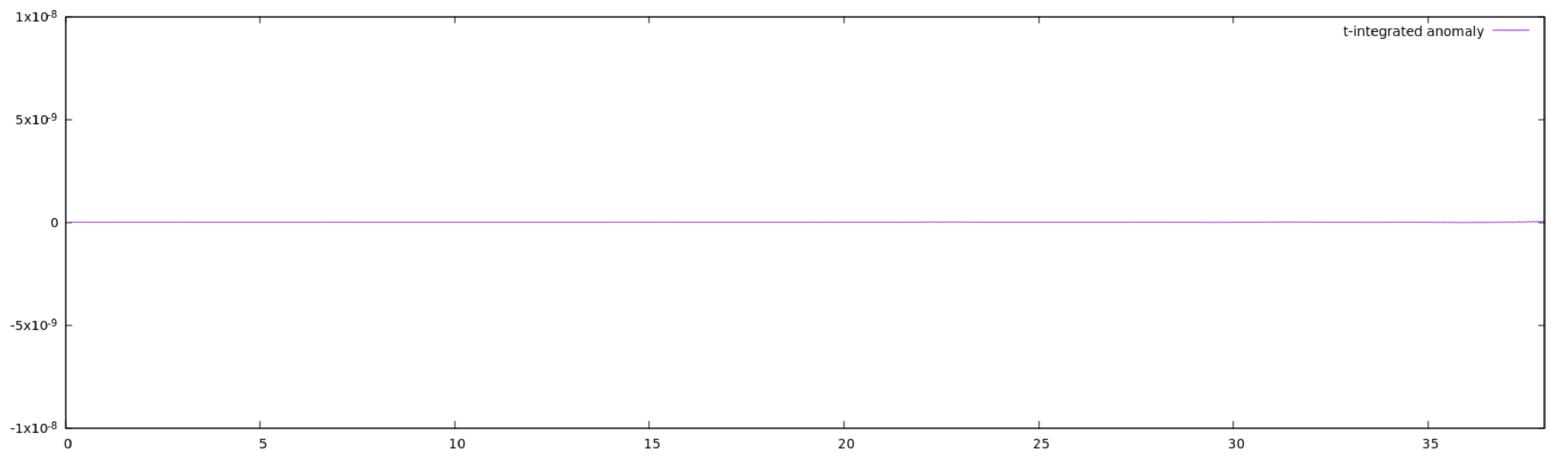}
\caption{Top Fig. shows the anomaly density $(w_x v_t V_x)\,$ plotted in $x-$coordinate for three successive times, $t_i =$ before collision (green), $t_c=$ collision (blue) and $t_f=$ after collision (red), for the 2-soliton of Fig. 1. Middle Fig. shows the plot of $ \int_{0}^{L} dx (w_x v_t V_x) \, \mbox{vs} \,\, t$ and  the bottom Fig. shows the $t-$integrated anomaly $\int_{t_o}^{t} dt \int_{0}^{L} dx (w_x v_t V_x)\,  \mbox{vs}\, \, t$. The last integral vanishes to within $10^{-9}$ precision.}
\label{fig5}
\end{figure}    

Let us consider some numerical results for the model (\ref{mrkdv}). Here we follow the finite difference method presented in \cite{jhep33}. In the Fig. 1 we plot the numerical result for 2-soliton collision with parameter values $\epsilon_1 = 1.2, \epsilon_2 = 0.9, \alpha = 4$ of the deformed KdV  model (\ref{mrkdv}). In Fig. 2 we show the results of the numerical simulations of the quantities related to the anomaly density in (\ref{vanish}): 1) The top Fig. shows the anomaly density $(w_x v_t V_x)\, \mbox{vs}\,x$ for three successive times. 2) The middle plot shows $\int_{0}^{L} dx (w_x v_t V_x)\, \mbox{vs}\,t$. 3) The bottom plot shows $\int_{t_o}^{t} dt \int_{0}^{L} dx (w_x v_t V_x)\, \mbox{vs}\,t$. Note that the bottom plot shows the vanishing of the space-time integrated anomaly, within numerical accuracy of the order of $10^{-9}$. So, it proves that (\ref{asym3a}) holds effectively with an accuracy on the order of $10^{-9}$.

Another approach to modify the KdV  equation in the context of quasi-integrability has been proposed by a direct deformation of its Hamiltonian in the Hamiltonian formulation of the model \cite{abhinav2}. The deformed KdV (\ref{dkdv2}), under $U\rightarrow \frac{1}{2} U, \, Y\rightarrow \frac{1}{2} Y $  becomes $U_t + 6 U U_x + U_{xxx}= Y_x$. So, the identification 
\br
\label{Y11}
Y \equiv 2 U^2 + \frac{2}{3} [\frac{\delta H_1}{\delta U} + U_{xx}],
\er
with $H_1$ being the deformed Hamiltonian, defines the deformed KdV in this approach. The undeformed Hamiltonian $H^{undef}_1$ is defined as
\br
\label{undh}
H^{undef}_1[U] = \int_{-\infty}^{+\infty} (\frac{1}{2} U_x^2-U^3).
\er 
Note that for the undeformed $H^{undef}_1$ (\ref{undh}) one has $Y=0$ in (\ref{Y11}). An explicit power modification of the nonlinear term in (\ref{undh}) as $U^{(3+3\epsilon)}$, with parameter $\epsilon$, has been discussed in \cite{abhinav2}. So, the Riccati-type approach, which is suitable for deforming the both nonlinear and dispersive terms, may also include the Hamiltonian deformation method to tackle quasi-integrable KdV models.
 
Furthermore, using the Riccati-type pseudo-potential approach, a linear system of equations corresponding to the deformed KdV model (\ref{dkdv2}) has been constructed. Within this linear formulation, an infinite set of non-local charges has also been identified \cite{jhep33, proc2}.

Moreover, given the widespread presence of KdV-type models across various areas of nonlinear science, it would be valuable to explore the significance and physical implications of the tower of asymptotically conserved charges identified above. For instance, a well-established connection exists between gravitation in three-dimensional space-times and two-dimensional integrable systems. Notably, the KdV-type and KdV-Gardner models have recently been recognized as governing the dynamics of the boundary degrees of freedom in General Relativity on AdS$_3$ (see, e.g., \cite{grav} and references therein). 
  
With regard to the asymptotically conserved charges, it is important to highlight that Ref. \cite{jhep33} presents an analytical-rather than purely numerical-proof of the quasi-integrability of the well-known non-integrable modified regularized long-wave (mRLW) theory, corresponding to the model defined in Eq. (\ref{mrkdv}) with $\epsilon_1=\epsilon_2=1$. Specifically, it has been demonstrated that the two-soliton solution of the mRLW theory, when expressed in a  ${\cal P}$-invariant form, exhibits quasi-integrable behavior in an analytical sense. This quasi-integrability applies to all anomalous charges associated with the infinite towers of quasi-conservation laws discussed earlier.

The dynamics of soliton-like configurations in quasi-integrable models remain largely unexplored. Key findings are as follows: (1) One-soliton sectors possess an infinite set of anomalous conserved charges, due to vanishing space–time integrals of anomaly densities.  (2) For two-soliton type solutions possessing the symmetry ${\cal P}$, the charges are asymptotically conserved. This means that these quantities vary in time during the collision process (and sometimes can vary quite a lot) of two one-solitons but return, in the distant future (after the collision), to the values they had in the remote past (before the collision). (3) For multi-soliton configurations with widely separated solitons, the anomalies also integrate to zero, with non-zero contributions localized to interaction regions.  (4) A sufficient condition for vanishing integrated anomalies is the presence of definite parity (even/odd) under the shifted parity and delayed time-reversal symmetry ${\cal P}$. Anomaly densities odd under this symmetry yield asymptotically conserved charges. (5) Multiple infinite towers of anomalous charges have been identified, expanding on earlier anomalous Lax results \cite{jhep1, jhep3, cnsns} with new contributions \cite{npb1, jhep33, paper1, paper2} for deformations of SG, KdV and NLS models based on Riccati-type formulations. (6) Certain deformed models exhibit subsets of infinite anomalous charges for soliton eigenstates invariant under shifted space-reflection symmetry ${\cal P}_s$\, (${\cal P}_s \subset {\cal P}$), as seen in deformed focusing/defocusing NLS models and deformed sine-Gordon models for two-soliton and breather solutions \cite{cnsns, jhep4, jhep5}. These results derive from combined analytical and numerical approaches.
  
In quasi-integrable models, the soliton boundary conditions studied in the literature fall into two main categories:
(1) {\bf Vanishing boundary conditions} at $x= \pm \infty$, associated with bright solitons in focusing modified NLS models \cite{jhep3, jhep5}, breather solutions in deformed sine-Gordon (SG) models \cite{jhep1, jhep2, cnsns}, and solitons in modified KdV models \cite{npb, jhep33}.
(2) {\bf Non-vanishing boundary conditions} at $x= \pm \infty$, relevant to dark solitons in defocusing modified NLS models \cite{jhep4, proc3}, kink solutions in deformed SG models \cite{jhep1, cnsns}, and kink solitons in deformed potential KdV (pKdV) models \cite{bjp1}.
Alternative boundary conditions—such as periodic or open—remain underexplored and warrant further investigation in the context of quasi-integrability.

\section{Conclusions and discussions}
\label{sec:discuss}
  
We have reviewed the quasi-integrability framework developed in 
\cite{npb,jhep33,proc2,paper1,proc3,paper2,bjp1} for deformations of the sine-Gordon (SG), nonlinear Schrödinger (NLS), and Korteweg–de Vries (KdV) integrable models, refining and unifying key arguments from these works. Our unified approach is based on the introduction of specific deformation fields, denoted as $r_j$ and $X$ in section \ref{sec:riccati}, into the Riccati-type pseudo-potential system associated with these integrable models. These deformation fields modify the underlying structure of the equations while preserving essential quasi-integrability properties, such as the existence of infinite towers of anomalous and non-local conservation laws. This formulation provides a systematic way to extend integrability concepts to deformed models, offering new insights into their soliton dynamics and conserved quantities.

It is demonstrated that deforming the Riccati-type pseudo-potential equations away from the SG model generates infinite towers of quasi-conservation laws for deformed sine-Gordon models of type \ref{sg1}. The first two conserved charges correspond to energy and momentum, while a related linear system enables the construction of an infinite set of non-local conservation laws. Additionally, through direct construction, further towers of quasi-conservation laws have been identified in \cite{npb1}.

In Sec. \ref{sec:dnls}, we applied the Riccati-type pseudo-potential approach to deformations of the AKNS model, obtaining the modified NLS (MNLS) through a specific reduction. This framework enabled the construction of infinite towers of quasi-conservation laws, related to the MNLS model. Our approach reproduced the NLS-type quasi-conserved charges derived via the anomalous zero-curvature method in \cite{jhep3}. It is introduced a dual Riccati-type pseudo-potential approach, revealing a new infinite set of quasi-conservation laws that encompass those obtained through a direct method from the equations of motion in \cite{paper1,proc3}.

In Section \ref{sec:dkdv}, we applied the pseudo-potential approach to KdV model deformations, demonstrating that modifying the Riccati-type pseudo-potential equations produces infinite towers of quasi-conservation laws for general KdV deformations (\ref{dkdv2}). This framework allowed us to construct an infinite set of quasi-conservation laws, recovering the KdV-type quasi-conservation laws derived via the anomalous zero-curvature method in \cite{npb1}. Additionally, we showed that the recently proposed Hamiltonian deformation approach to quasi-integrability \cite{abhinav2} aligns with our general pseudo-potential framework.

Partial differential equations have been interpreted as infinite-dimensional manifolds, leading to the introduction of differential coverings. These coverings have been employed to construct various structures, such as Lax pairs and B\"acklund transformations (see e.g., Refs. \cite{krasil1, krasil2, igonin, morozov} ). Notably, an auto-B\"acklund transformation corresponds to an automorphism of the covering. Therefore, it would be valuable to explore the properties of the Riccati-type System (\ref{riccati11})-(\ref{riccati22}) as a form of differential covering for the corresponding modified integrable system.

An interesting direction for future research is the dynamics of kinks and bound-state solutions in soliton-fermion quasi-integrable models, such as the deformed SUSY-SG \cite{epl}. Notably, the integrable affine Toda model coupled to fermions (ATM) was recently studied in \cite{jhep22}, revealing strong-weak duality in soliton-fermion sectors. This investigation could also have significant implications for one-dimensional topological superconductors \cite{eneias}, where understanding the impact of interactions on the shape and lifetime of bound states is crucial. In particular, recent proposals to identify Majorana modes through local integrals of motion in interacting systems \cite{maska} highlight the relevance of this study. Therefore, further analysis of quasi-integrable versions of the ATM model is warranted.

Furthermore, the relationship between quasi-integrability and non-Hermiticity deserves further investigation. The non-Hermitian affine Toda model coupled to fermions was studied in \cite{jhep24} using soliton theory techniques, exploring pseudo-chiral and pseudo-Hermitian symmetries. The interplay between non-Hermiticity, quasi-integrability, nonlinearity, and topology could play a crucial role in the formation and dynamics of exotic quantum states, with potential applications in topological quantum computation. Additionally, studying quasi-integrability and non-Hermitian solitons may offer new insights into the theory of topological phases of matter.

Recent studies have explored non-Hermitian extensions of certain integrable models. This direction presents a compelling avenue for future investigation within the framework of quasi-integrability. In principle, the Riccati-type pseudo-potential method is applicable to a broad class of integrable systems, provided that their standard Riccati formulation is established. However, the analysis of novel quasi-integrable features emerging from such non-Hermitian deformations lies beyond the scope of the present review.

Finally, a deeper investigation is needed into the physical nature and dynamics of deformed solitons, particularly their behavior in collision regions and the underlying mechanisms governing anomaly cancellation to restore exact conservation laws, as suggested in \cite{paper1,proc3, bjp1}. Understanding how these solitons interact and evolve under deformations could provide further insights into quasi-integrability and its broader implications. Additionally, exploring potential applications of these findings in nonlinear physics, such as in Bose–Einstein condensation \cite{frantzeskakis1,frantzeskakis2}, superconductivity\cite{sc1, tanaka1}, soliton turbulence \cite{gauss1}-\cite{turbu2}, and soliton gas \cite{prlgas} remains an open avenue for research.

\end{document}